\documentclass[]{article}
\usepackage[dvipsnames]{xcolor}
\usepackage{graphicx}

\usepackage[version=4]{mhchem}
\usepackage{amsmath}
\usepackage{amssymb}
\usepackage{color}
\usepackage{overcite}
\usepackage[font=small,labelfont=bf]{caption}

\newcommand{\R}{\mathbb{R}}

\usepackage[framemethod=tikz]{mdframed}
\newmdtheoremenv[ 
hidealllines=true,
leftline=true,
innertopmargin=0pt,
innerbottommargin=0pt,
linewidth=4pt,
linecolor=gray!40,
innerrightmargin=0pt,
innertopmargin=-6pt,
]{examplei}{Remark}

\title{\bf Quantum chemical roots of machine-learning molecular similarity descriptors}
\date{September 11, 2022}
\author{\vspace{0.3cm}
        Stefan Gugler and Markus Reiher\footnote{Corresponding author; e-mail: markus.reiher@phys.chem.ethz.ch}\\
        \textit{ETH Z\"urich, Laboratorium f\"ur Physikalische Chemie,}\\
        \textit{Vladimir-Prelog-Weg 2, 8093 Z\"urich, Switzerland}\\[2ex]
}

\begin{document}
	\maketitle
	
	\renewcommand*{\thefootnote}{\fnsymbol{footnote}}
	
	\vspace*{0.2cm}

	\begin{abstract}
In this work, we explore the quantum chemical foundations of descriptors for molecular
similarity. Such descriptors are key for traversing chemical compound space with machine learning. 
Our focus is on the Coulomb
matrix and on the smooth overlap of atomic positions (SOAP). We adopt a basic framework that allows us to
connect both descriptors to electronic structure theory. 
This framework enables us to then define two new descriptors that are more closely related to electronic structure
theory, which we call Coulomb lists and smooth overlap of electron densities (SOED). By investigating their usefulness as
molecular similarity descriptors, we gain new insights into how and why CM and SOAP
work. Moreover, Coulomb lists avoid the somewhat mysterious diagonalization step of the
CM and might provide a direct means to extract subsystem
        information that can be compared across Born--Oppenheimer
surfaces of varying dimension. For the electron density we derive the necessary formalism to create the
SOED measure in close analogy to SOAP. Because this formalism is more involved than that of SOAP,
we review the essential theory, as well as introduce a set of approximations that eventually allow us
to work with SOED in terms of the same implementation available for the evaluation of SOAP.
{We focus our analysis on elementary reaction steps, where
transition state structures are more similar to either reactant or
product structures than the latter two are with respect to one another.
The prediction of electronic energies of transition state structures
can, however, be more difficult than that of stable intermediates due
to multi-configurational effects. The question arises to what extent
molecular similarity descriptors rooted in electronic structure theory
can resolve these intricate effects.}
	\end{abstract}

	\section{Introduction}
	
	A molecular descriptor---in the machine-learning literature also known
	as feature\cite{bishop2006}---is a
	representation of a molecule in terms of a computer readable vector.
        Molecular descriptors can be compared in order to assess the similarity of molecules of different 
        composition and configuration.
	A similarity measure is a mathematical metric, that is,
	a function that measures the distance between two points 
        in descriptor space.
	The closer the two points are, the more similar they are.

	A kernel function generalizes this notion.\cite{mercer1909}
	Examples are Gaussian kernels or linear kernels.\cite{bishop2006,hastie2001}
	They do not bear physical meaning per se but assess whether two points
	are close together according to the measure.
	Often, they are chosen for mathematical convenience (such as 
	radial basis functions based on Gaussians\cite{rasmussen2006}),
	but they can also be loaded with physical interpretation such as the
	smooth overlap of atomic positions (SOAP)\cite{bartok2013}. 
	Even though the notions of features and kernels are distinct,
	they are sometimes treated as practically the same, because
	the physical and chemical properties of systems to be represented are encoded in both the
	kernel and the feature.
	
	There exists a multitude of molecular descriptors, many
	originating from the field of cheminformatics:
	one of the most comprehensive overviews was provided by Todeschini and
	Consonni\cite{todeschini2009}, which, however, does not cover post-2012
	descriptors, that is, those that have gained traction within the
        recent revival of machine learning and artificial intelligence in chemistry.
	Todeschini and Consonni\cite{todeschini2009} 
        classify the descriptors according to the theory from which they
	are derived: 
	graph theory, discrete mathematics, physical chemistry, information theory,
	quantum chemistry, organic chemistry, differential topology, and algebraic
	topology.
	On top of that, they distinguish how they are processed,
	namely, by statistics, chemometrics, or cheminformatics.
	The bibliography of their review covers the period between 1741 and 2008 with
	about 6400 references and 3300 descriptors listed.\cite{todeschini2009}
	
	With the advent of modern machine learning in chemistry, several
	groups have developed descriptors and methods that are better tailored
	to a particular machine-learning method and harness the latest computational developments
	more efficiently.
	
  {The group of von Lilienfeld proposed various descriptors, most of which
  can be described as two- or many-body expansions such as the
        atom-in-molecule-based descriptor called ``amons''\cite{huang2020} or
        FCHL\cite{christensen2020} (named after the authors). In collaboration
        with the groups of M\"{u}ller and Tkatchenko, they
        developed\cite{rupp2012} and assessed\cite{hansen2013,schrier2020}
        Coulomb matrices (CMs), their eigenvalues, or multiple CMs per
        molecule, as well as the bag of bonds (BoB) descriptor.\cite{hansen2015}
        Huang and von Lilienfeld studied the uniqueness
        of some of their descriptors.\cite{huang2016}}

  {A variation in the CM that avoids the permutation issue is the
        BoB descriptor.\cite{hansen2015} All off-diagonal
        elements of the CM are collected and grouped (``bagged'')
        according the corresponding element tuples which is interpreted as bond
        ``type'' (such as the tuples (H,H) and (C,H)). Each group of Coulomb
        matrix elements is then sorted by magnitude and padded with zeros if
        other molecules in the data set have more elements in a given bag.
        Hence, all bags have the same length. The feature vector is the
        concatenation of all bags. This idea has been developed into a
        hierarchy of features including more and more bodies, collectively
        called ``BA-representations'' (bonds, angles, torsions, etc.) to hint
        at the many-body expansion character.\cite{huang2016} It commonly
        performs slightly better than the CM
        itself.\cite{hansen2015,collins2018,pronobis2018}}

	To establish size-intensive descriptors, Collins studied descriptors for
	machine learning and how to encode bonds.\cite{collins2018}
        The molecular-structure-based descriptor SOAP
    by Cs\'{a}nyi and co-workers has experienced continuous development and is among the most successful ones.\cite{bartok2013,de2016,willatt2019}
	Ceriotti and co-workers have developed and analyzed physics-inspired molecular representations\cite{musil2021,bartok2017,ceriotti2021} and carried out considerable
	work in unifying the landscape of descriptors.\cite{pozdnyakov2020b,musil2019a,musil2021,goscinski2021}
  {Various groups have investigated ways to incorporate the electron density
  into machine learning procedures.\cite{grisafi2019b,fabrizio2020,willatt2019,tang2018,geidl2015}}

  {Other descriptors use topological information\cite{balaban1982,modee2021}
        or a graph-based
        representation,\cite{balaban1985,mohapatra2021,mcdonagh2018} both of
        which have deep roots in cheminformatics and find applications in
        machine learning for chemistry. Working a set of empirical topological
        descriptors\cite{janet2017a} into autocorrelation
        functions\cite{broto1984} has been applied to transition metal
        complexes by Janet and Kulik.\cite{janet2017c} Neural-network
        representations have been studied by Behler and Parrinello with
        symmetry function to represent local
        environments\cite{behler2007,behler2011a} and by Aspuru-Guzik and
        co-workers, where the neural network operates directly on the molecular
        graph.\cite{duvenaud2015} Instead of finding an adequate
        representation, end-to-end neural networks were shown to learn the
        appropriate representation on their
        own.\cite{schutt2017b,schutt2018b,lubbers2018,unke2019,scarselli2009}}

    {This work is in the regime of low amounts of data as we focus, for the sake of a detailed analysis, on a specific 
similarity paradox in reaction chemistry.}

Molecular similarity descriptors such as CM and SOAP have been used in
machine-learning applications as ``rulers'' to assign a degree
of similarity to two different structures. Typically, the structures to be compared can differ significantly and may be taken
from across chemical space. However, if we consider the opposite case, namely, structures that are clearly related through
an elementary reaction step, then we arrive at a somewhat paradoxical situation: while reactant and product are obviously 
rather similar molecules by construction (as they are related by an elementary reaction step), the connecting
transition state (TS) structure is even more similar to either product or reactant. Yet, it is well known in electronic
structure theory that TS structures (which represent activated molecules that typically exhibit 
one or two stretched chemical bonds)
present a very different electron correlation problem compared to the stable reactant or product structures.
Hence, this example presents a well-defined situation, in which molecular structure similarity seems to be
insufficient to also judge electronic similarity, which governs the molecular properties. As a result, 
a molecular similarity measure might rightfully determine a TS structure to be more similar to
either side of the reaction arrow but at the same time miss the fact that its electronic structure will be rather
different (as measured, for instance, in terms of electron correlation diagnostics). Another measure might even determine all 
three structures to be basically identical, which will then ignore all fine-grained differentiation that quantum
chemical methods attributes to them.

  {We note that machine learning has also been applied to TSs\cite{abdelfatah2019,lemm2021,jackson2021,zhang2021,pozun2012,kayala2011,chen2022}
  compared to stable intermediates, but more can be expected in the future in
  view of the key role TSs play in chemistry. Closely related is
  the machine learning of reaction barriers.\cite{singh2019, simm2019a,
  heinen2021, yang2020, friederich2020, takahashi2018}}
It is therefore important to better understand to what degree descriptors of molecular similarity can 
differentiate between stable intermediates (i.e., local minima on the Born--Oppenheimer potential energy surface) and
TSs (i.e., first-order saddle points on that surface).

{Kulik and co-workers studied the applicability of the eigenvalues of the CM for transition metal complexes.\cite{janet2017a,janet2017c, nandy2021} Compared to their autocorrelation functions proposed as descriptors, the CM eigenvalue performed poorly. They found that electronic properties such as spin-splitting energies are not represented well and show a large dependence on  molecular size. However, despite these practical difficulties
the CM should be applicable to the whole range of the periodic table 
due to its connection to the first principles of quantum mechanics. 
Its explicit dependence on the external potential (nuclear charges and nuclear positions)
and hence its direct connection to electronic structure theory should allow for the differentiation of atoms in molecules and, at the same time, for the identification of isoelectronic features. 
  Moreover, we will later see that SOAP can also be rooted in first-principles electronic structure theory and generality 
across the periodic table may be expected. In fact, it has been used for
various materials\cite{bartok2017, deringer2021} and elements such as
boron\cite{deringer2018} or silicon\cite{deringer2018a}, which also indicates
general applicability. Along these lines, arguments for general applicability
can also be made for the two descriptors introduced in this work: Coulomb lists
(CL) and smooth overlap of electron densities (SOED).
Whether or not this can be exploited in practice remains to be a subject of
future studies as it is far beyond the scope of the present work.}

  {In this work, we consider the physical foundations of the CM
  and SOAP from the point of view of electronic structure theory. The selection
  of these two out of the many available descriptors is driven by the
  possibility to connect to electronic structure theory and justified by their
  popularity. Moreover, we aim at non-neural, low-data regime descriptors that
  have been influential in the current surge of machine learning for
  chemistry.
  In addition, they are commonly taught in introductory courses to machine
  learning for chemistry and are available in machine learning software for
  chemical applications.
  \cite{himanen2020,dral2019,collins2018,haghighatlari2019,khatib2020,musil2021b,caro2019}}
  {We first introduce three example reactions for which we will compare the
  different descriptors, unlike in a machine learning study, where a large data
  set is used to assess the predictive power of a descriptor.}
        Then, we study a general expression for the electronic energy to which we
        want to relate the measures in order to establish a relation between a similarity measure and this key
        quantity of electronic structure theory. Afterwards, we first consider the relation of the electronic energy
        and the CM, which will also lead us to introduce CLs as a descriptor. Then, we turn to
        such a relation for SOAP, which will also lead us to the introduction of a new descriptor, namely the
        SOED.

	\section{Elementary Reaction Step Example}

By contrast to a typical machine learning study, we will, in this first step, not consider vast amounts of data, but
instead make an attempt to understand how CM and SOAP operate at the level of a single elementary reaction step. In order
to be able to later extend this work to a large data set, we chose our
(generic) example (reaction 1) in such a way that it is on the same Born--Oppenheimer
surface as the QM9 reference data set of von Lilienfeld and co-workers \cite{ramakrishnan2014}.
	
\begin{figure}
		\centering
		\includegraphics[width=0.7\textwidth]{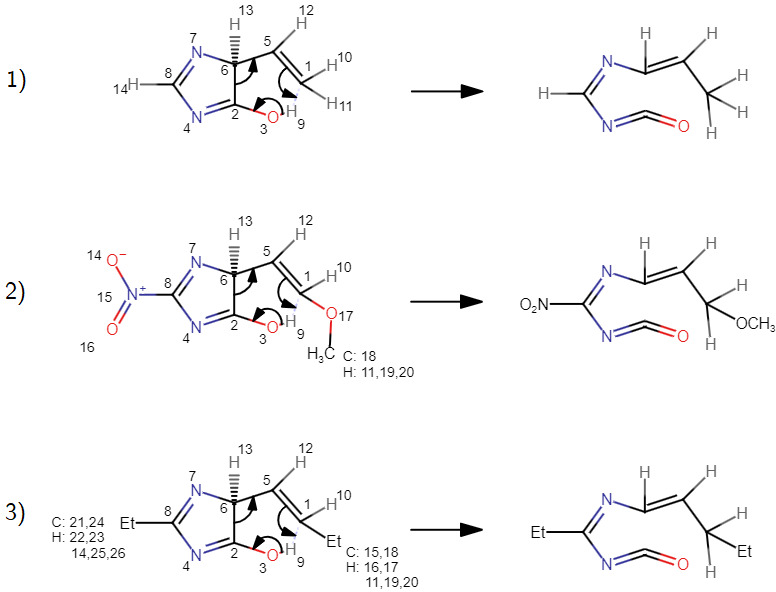}
		\caption{Reactions for which an elementary step relates reactant and product through a TS structure.
			In (the generic) reaction 1, the nucleophilic double bond between \ce{C1} and \ce{C5} abstracts the proton
			of the hydroxyl group, \ce{H9}. The \ce{C2}-\ce{C6} bond breaks and, hence, opens up the 
			imidazole ring, yielding an isocyanate. 
			Reaction 2 and 3 show analogous reactions with a push--pull system
            (in 2) and two ethyl substituents (in 3), respectively.
		}
		\label{fig:rxn}
	\end{figure}

	We will base the numerical part of this work on the reactions shown in Figure \ref{fig:rxn}: 
	the nucleophilic double bond between \ce{C1} and \ce{C5} abstracts the proton 
	of the hydroxyl group, \ce{H9}, which results in a change of the aromatic system as the imidazole
	ring opens up.
	An isocyanate group is created.
	In this generic reaction 1, both electronic and structural changes are present.
	This allows us to probe to what extent a descriptor can account for
	electronic changes in terms of its numerical values.

Analogously to reaction 1, reactions 2 and 3 are chosen as proton abstraction reactions of the same type but with
modulating substituents. 
In the case of reaction 2, we introduced a push--pull-type $\pi$-system structure 
to replace two hydrogen atoms of the original reaction. The nitro group functions
as an electron acceptor, and the methoxy group may be considered as the electron donor. 
Accordingly, this system poses a challenge for a descriptor. 
In the case of reaction 3, two ethyl groups replace the same two hydrogen atoms to probe the effect 
of atoms that are farther away from the reactive site.
Hence, the three reactions serve the purpose of studying descriptor transferability because the 
        locality of the descriptor should highlight whether the generic reaction 1 can be recovered
        in reactions 2 and 3.

	\section{Electronic Energy in Terms of Nuclear Contributions}\label{sec:components}
	
To connect a descriptor of molecular similarity, which refers to an atom-resolved molecular structure (usually 
in terms of its nuclear coordinates), with
the expression for the electronic energy, we first need to rewrite this energy in terms of nuclear contributions.
Although every electronic structure model affords a different expression for the electronic energy, a common
expression can be formulated in terms of one- and two-body reduced density matrices.
If we focus on the nuclear contributions in such an expression, we may write the electronic energy
	in Hartree atomic units as:
	
	\begin{equation}
		E_\text{el} = 
		\sum_{I>J}^{M} \frac{Z_{I}
			Z_{J}}{\left|\mathbf{R}_{I}-\mathbf{R}_{J}\right|}
		-\sum_{i,J}^{N,M} n_i
		\left\langle\phi_{i}\left|
		\frac{Z_{J}}{\left|\mathbf{R}_{J}-\mathbf{r}\right|}
		\right| \phi_{i}\right\rangle
		+E_\text{rest}[\{\phi_i\}]\, ,
		\label{eq:el-ene}
	\end{equation}
	
	where $I$ and $J$ are the indices running over the $M$ nuclei with their respective
	nuclear charges, $Z_I$ and $Z_J$, and coordinates, $\mathbf{R}_I$ and $\mathbf{R}_J$.
	The first sum refers to the Coulomb repulsion of all nuclei, which is independent of the electronic structure model.

	The second term delivers the potential energy for the interaction with the external potential and all other contributions are kept hidden in the third one.
        For the sake of simplicity, we have introduced a general occupation number $n_i$, which can be easily generalized to
        the doubly indexed one-body reduced density matrix.
        One may simply choose the $n_i$ to be equal to 1 as in unrestricted
        Hartree--Fock or Kohn--Sham theory.
        The index $i$ then runs over $N$ spin orbitals $\phi_i$ from which the electronic wave function is constructed.
        Finally, ${\bf r}$ is the coordinate of an electron.
	
	$E_\text{rest}[\{\phi_i\}]$ denotes the remaining contributions to the electronic energy, namely the expectation values for
	total kinetic energy, $\langle T\rangle$, and the electron--electron interaction, $\langle V_\text{ee} \rangle$, 
	
	\begin{equation}
		E_\text{rest} = \langle T \rangle + \langle V_\text{ee} \rangle \, ,
	\end{equation}
        which all depend on the wave function and, hence, on all orbitals $\{\phi_i\}$.
	If molecular orbitals (MOs) are expanded into a set of $m$ basis functions $\chi_\mu$,
	\begin{equation}
		\phi_i = \sum_{\mu=1}^{m} c_{\mu}^{(i)} \chi_{\mu} \, ,
		\label{eqn:simple-expansion}
	\end{equation}
	their corresponding MO coefficients $c_\mu^{(i)}$ will enter the energy expression
	\begin{equation}
		E_\text{el} = 
		\sum_{I>J}^{M} \frac{Z_{I} Z_{J}}{\left|\mathbf{R}_{I}-\mathbf{R}_{J}\right|}
		-\sum_{i,J}^{N,M} n_i
		\sum^m_{\mu \nu} c_{\mu}^{(i)} c_{\nu}^{(i)}
		\left\langle\chi_{\mu}\left|
		\frac{Z_{J}}{\left|\mathbf{R}_{J}-\mathbf{r}\right|}
		\right| \chi_{\nu}\right\rangle
		+E_\text{rest}[\{c_{\mu}^{(i)} \}] \,  .
	\end{equation}
Note that we assumed the basis functions to be real and therefore avoided a denotation for complex conjugation;
however, this restriction can be easily lifted.
	
	We re-write the electron--nucleus attraction potential-energy integrals over basis functions $\chi_\mu$
	into a matrix, $\mathbf{V}^{(J)}$, with elements
	
	\begin{equation}
		V_{\mu \nu}^{(J)} = \left\langle\chi_{\mu}\left|\frac{Z_{J}}{\mid \mathbf{R}_{J}-\mathbf{r}\vert}\right| \chi_{\nu}\right\rangle
	\end{equation}
	and the MO coefficients
	into the vector, $\mathbf{c}^{(i)}$, obtaining
	
	\begin{align}
		E_\text{el}
		=& \sum_{I>J}^{M} \frac{Z_{I} Z_{J}}{\left|\mathbf{R}_{I}-\mathbf{R}_{J}\right|}
		-\sum_{i,J}^{N,M} n_i
		\mathbf{c}^{(i)} 
		\mathbf{V}^{(J)}
		\mathbf{c}^{(i)}
		+E_\text{rest}[\{c_{\mu}^{(i)} \}] .
		\label{eqn:eel}
	\end{align}

	Obviously, for a close connection of the quantum chemical foundations to a molecular similarity descriptor of a machine learning model, we may require that a
        descriptor should be based on Cartesian coordinates of all atomic nuclei, $\{\mathbf{R}_{I}\}$, 
        of a molecule because these define a molecular
        structure to which the Born-Oppenheimer approximation assigns an electronic energy. CM
        and SOAP fulfill this requirement (see below). Moreover, according to the electronic energy expression, we
        require that the descriptor depends on the nuclear
        charge numbers, which, together with the nuclear coordinates, defines the external potential. 
        The external potential and the number of electrons in the system contain all information to formulate the
        electronic Hamiltonian and, hence, all information to solve for the electronic energy (E. Bright Wilson argument \cite{handy1994}).
        The CM fulfills this requirement by construction, but it should be noted that it involves a diagonalization 
        step that changes the information encoded in a rather non-transparent way. 
        However, the standard formulation of SOAP considers molecular structure through fuzzy atoms, where all atoms are
        considered equal (see below). As such, nuclear coordinates enter the procedure, but their type (in terms of the
        nuclear charge number) is usually not resolved. 
        According to the first Hohenberg--Kohn theorem \cite{hohenberg1964},
        which is also taken as the basis of density functional theory (DFT), 
        there exists a one-to-one correspondence between the external potential and the electron density.
Because SOAP constructs a density distribution of molecular structure, one may wonder whether a relation to the
electron density (and hence to the electronic energy) can be established. We will consider these matters later on in
this work and now first turn to the CM due to its obvious link to the external potential.

	\section{CM and CL}
	The elements of the CM\cite{rupp2012} are defined as
	\begin{equation}
		C_{IJ} = 
		\begin{cases}
			\dfrac{Z_I^{2.4}}{2} \quad &I=J \\
			\dfrac{Z_I Z_J}{|\mathbf{R}_I - \mathbf{R}_J| } \quad &I\neq J \, .
		\end{cases}
	\end{equation}
	The diagonal in the CM is sometimes referred to as the
	``self-interaction term''\cite{hansen2013}, 
	even though there is no basis in classical electrodynamics for a self-interaction of a
	point charge.
	A pragmatic justification is that the diagonal term conveys information
	about the identity of the elements.
	The original publication\cite{rupp2012} states that the
	``diagonal elements encode a polynomial fit of atomic energies to nuclear charge''. 
	Because the authors wanted to predict atomization energies, this
	diagonal brings relevant information into the problem from a
	priori knowledge but makes the descriptor less general and less
	interpretable. 
Yet another interpretation\cite{ramakrishnan2015b} is to link the diagonal terms to the total potential energy of a neutral atom in the Thomas--Fermi model, which is $E_\mathrm{TF} = -0.7687 Z^{7/3}$.\cite{parr1995a}
	
	Molecules of different sizes, $M$, will be padded with rows and columns of zeros in
	their CM representation to match the size of the largest molecule to be compared.
	The CM itself cannot be used as a descriptor for molecules
	of different sizes, because it is not
	permutationally invariant (exchange of rows and columns change the
	descriptor but physically, the order of the atoms in the molecule does
	not matter) and different information would be stored in different dimensions
	of the matrix.
	
	Three remedies have been proposed\cite{hansen2013} 
	to transform the CM into a permutationally invariant descriptor.
	The simplest one, which we are going to analyze here, is the eigenspectrum:
	calculating the eigenvalues,
	$\lambda_i$, and sorting them such that $\vert \lambda_i \vert \geq
	\vert \lambda_{i+1} \vert$. This is the original recipe proposed by Rupp {\it et al.} in
	2012.\cite{rupp2012} For $M$ atoms, this method
    reduces the dimensionality from $M^2$ degrees of freedom to
	only $M$.
	The second option to make the CM permutationally invariant is the sorted CM:
	the rows (or equivalently, the columns)
	are sorted by their Euclidean norm, such that
	$\vert \vert C_i \vert\vert_2 \geq \vert \vert C_{i+1}
	\vert\vert_2$. This leads to an overdetermined system, as the
	dimensionality is $M^2$ now and may produce to non-smooth changes
	in the sorting even for small changes in the coordinates.
	The third approach is to represent each system by a set of $n$ sorted 
	CMs, each injected with Gaussian noise to vastly augment dimensionality.

	\subsection{Atomic Descriptors $\mathcal{F}_{\text{n}}^{(J)}$ and $\mathcal{F}_{\text{e}}^{(J)}$}
	
	We now split Eq.\ (\ref{eqn:eel}) into atomic contributions by moving the sum over the
	nuclei in front of the expression
	
	\begin{equation}
		E_\text{el}
		= \sum^M_J \left[
		\mathcal{F}^{(J)}_\text{n} -
		\mathcal{F}^{(J)}_\text{e}
		\right]
		+E_\text{rest}
	\end{equation}
	
	with
	
	\begin{equation}
		\label{eqn:fnn}
		\mathcal{F}^{(J)}_\text{n} = \frac{1}{2} 
		\sum_{I \atop I\neq J}^{M} \frac{Z_{I} Z_{J}}{\left|\mathbf{R}_{I}-\mathbf{R}_{J}\right|}
	\end{equation}
	
	and
	
	\begin{equation}
		\label{eqn:fne}
		\mathcal{F}^{(J)}_\text{e} =
		\sum_{i}^{N} n_i
		\mathbf{c}^{(i)} 
		\mathbf{V}^{(J)}
		\mathbf{c}^{(i)}
	\end{equation}
(recall that all $n_i$ are equal to one for unrestricted Hartree--Fock theory and unrestricted Kohn--Sham theory).
	
\begin{figure}
		\centering
		\includegraphics[width=0.8\textwidth]{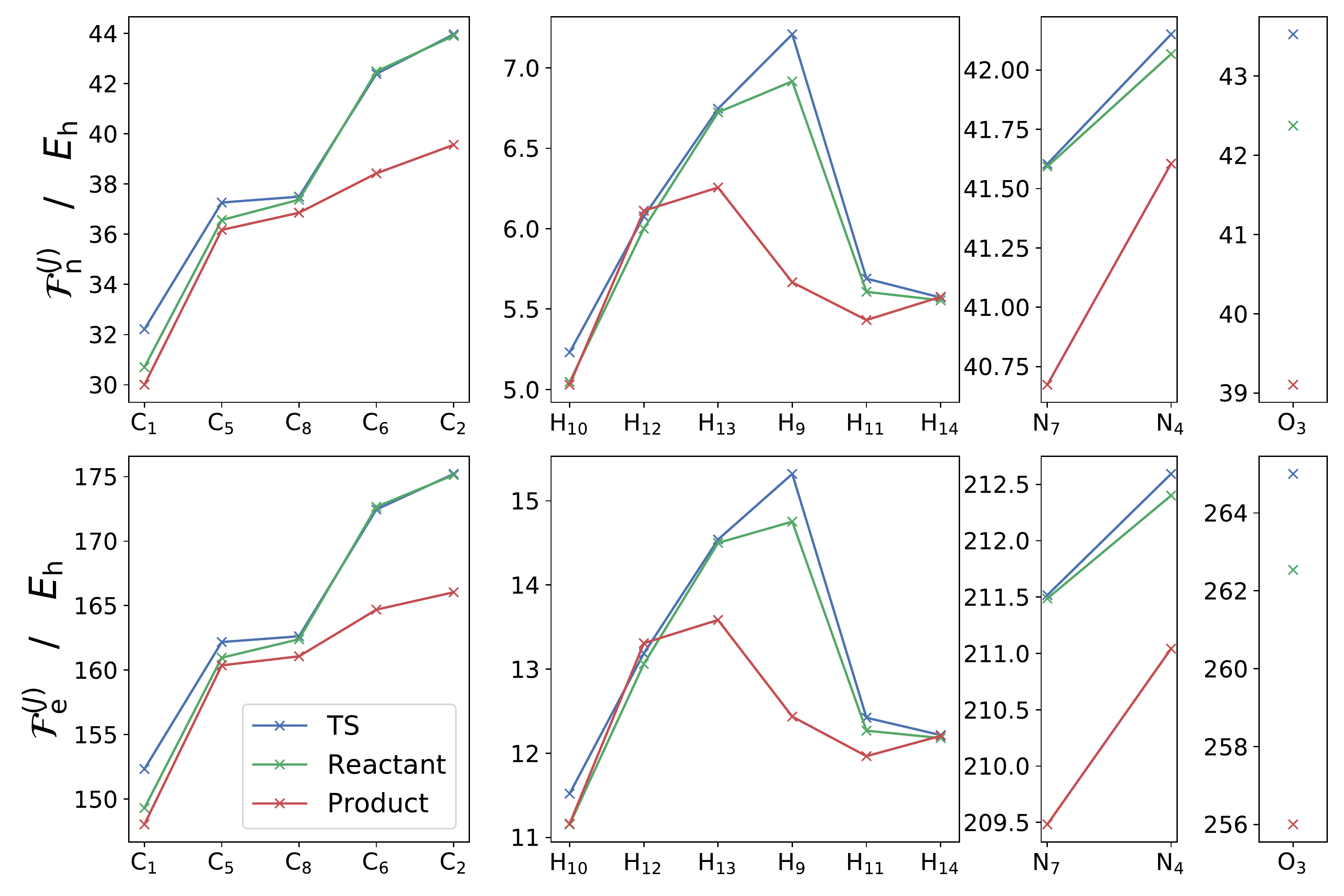}
		\caption{For reaction 1 of Figure \ref{fig:rxn}:
			(top) nuclear repulsion features, $\mathcal{F}^{(J)}_{\text{n}}$, 
            and
			(bottom) external potential features, $\mathcal{F}^{(J)}_{\text{e}}$,
			in hartree ($E_\mathrm{h}$)
 (TS denoted in blue and the reactant and product in green and red, respectively).
			The four subplots are sorted by element and to match the
            ordering of Figure 4.
		}
		\label{fig:relation_fnn_fne}
	\end{figure}

	Figure \ref{fig:relation_fnn_fne} depicts the features $\mathcal{F}_{\text{e}}^{(J)}$ and $\mathcal{F}_{\text{n}}^{(J)}$ 
	obtained for reaction 1 of Figure \ref{fig:rxn}.
	In each subplot of Figure \ref{fig:relation_fnn_fne}, we separated the elements due to the large difference in scale.
	In (a), the external potential features, $\mathcal{F}_{\text{e}}^{(J)}$, are shown, and
	in (b), the nuclear repulsion features, $\mathcal{F}_{\text{n}}^{(J)}$, are shown.
	The same plot scaled by the respective nuclear charge can be found in the Supporting Information.
	The scaling changes only the relationship between different elements but not within one group of elements.
	
	We see how the reaction details can be recovered in the plots of Figure \ref{fig:relation_fnn_fne}.
	For instance, \ce{C6} and \ce{C2} (and similarly \ce{H13}) show a big drop from the TS structure
	(blue line) to the product (red line), which monitors that the bond between these
	two atoms is broken.
	Similarly, \ce{H9} features the biggest relative drop toward the product,
	which is due to the H shift observed in the reaction.
	Only for \ce{H12} is the product feature higher in energy than those of the TS structure and
	the reactant.
	The fact that the reactant and the TS structure features are close together
	while the product features are separated hints at a possible early TS
	structure, which resembles the reactant rather than the product according to Hammond's postulate.\cite{meany2001}
		
\begin{figure} 
		\centering
		\includegraphics[width=1\textwidth]{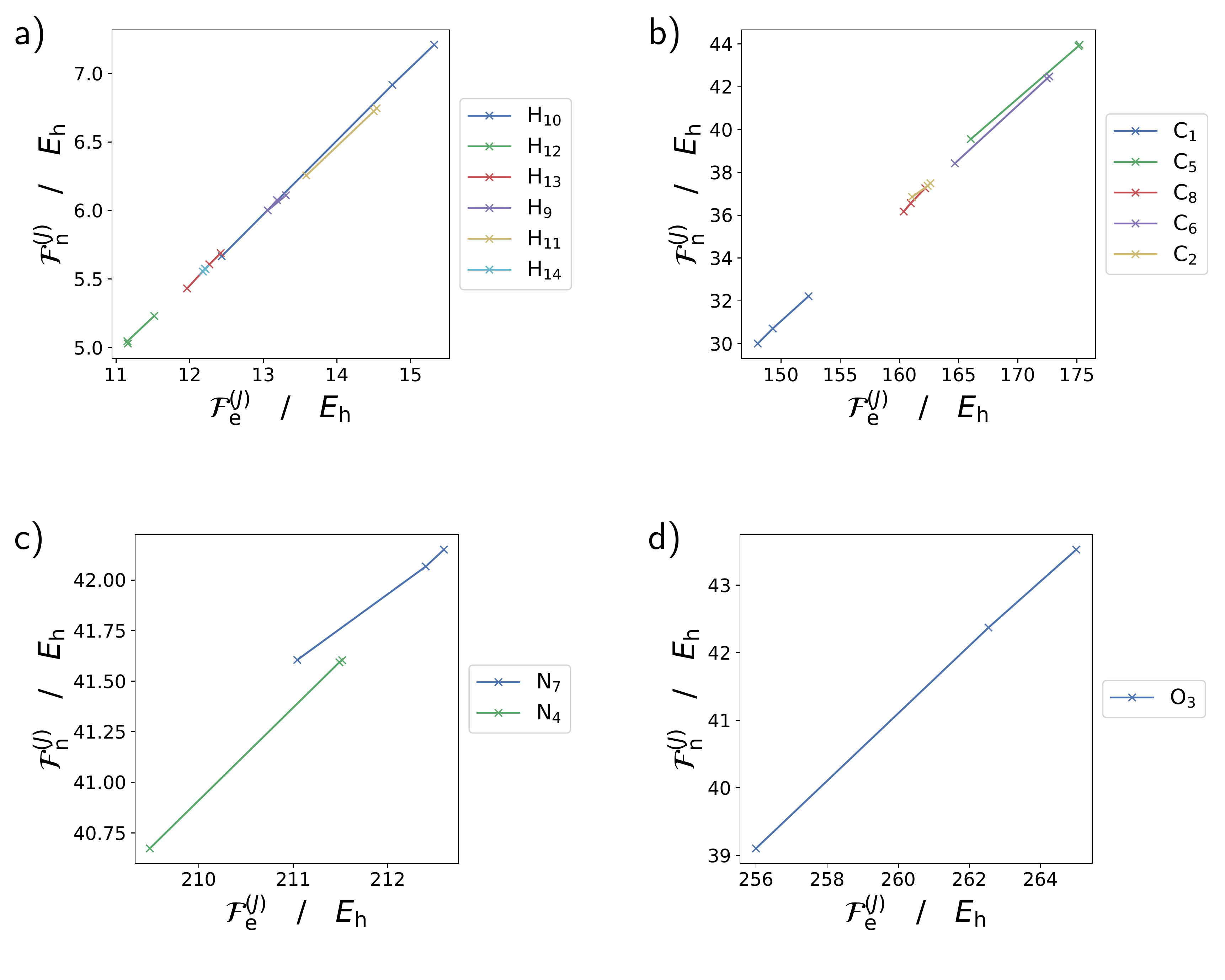}
		\caption{
			For each element in reaction 1 from Figure \ref{fig:rxn}, hydrogen, carbon, nitrogen, and oxygen in (a-d), we
			show the correlation between the external potential features, $\mathcal{F}_{\text{e}}^{(J)}$,
			and the nuclear repulsion features, $\mathcal{F}_{\text{n}}^{(J)}$.
			The connection between three points is over the course of the reaction from reactant to TS structure to product.
		}
		\label{fig:correlation_fnn_fne}
	\end{figure}

	To further elucidate the correlation, we plot $\mathcal{F}_{\text{n}}^{(J)}$ and $\mathcal{F}_{\text{e}}^{(J)}$
	against each other for each element in Figure \ref{fig:correlation_fnn_fne}.
	Apart from the scaling, the trends are very similar: Not only are all elements of the same
	type almost linearly correlated, but also the trend holds over the course of the reaction (connected points).
	Thus, apart from small deviations, the nuclear features encode very similar information to the external potential.
	Because the latter is much more expensive to calculate, it is a good trade-off to use the nuclear features,
	which are much more cost effective.
	This is evidence that at least approximately, only one of the two features can be used without great loss
	of accuracy. 
	Because the nuclear features are far more efficient to calculate, as they do not depend on a converged SCF calculation,
	we will solely consider $\mathcal{F}_{\text{n}}^{(J)}$ in what follows.
	
\begin{figure}
		\centering
		\includegraphics[width=1\textwidth]{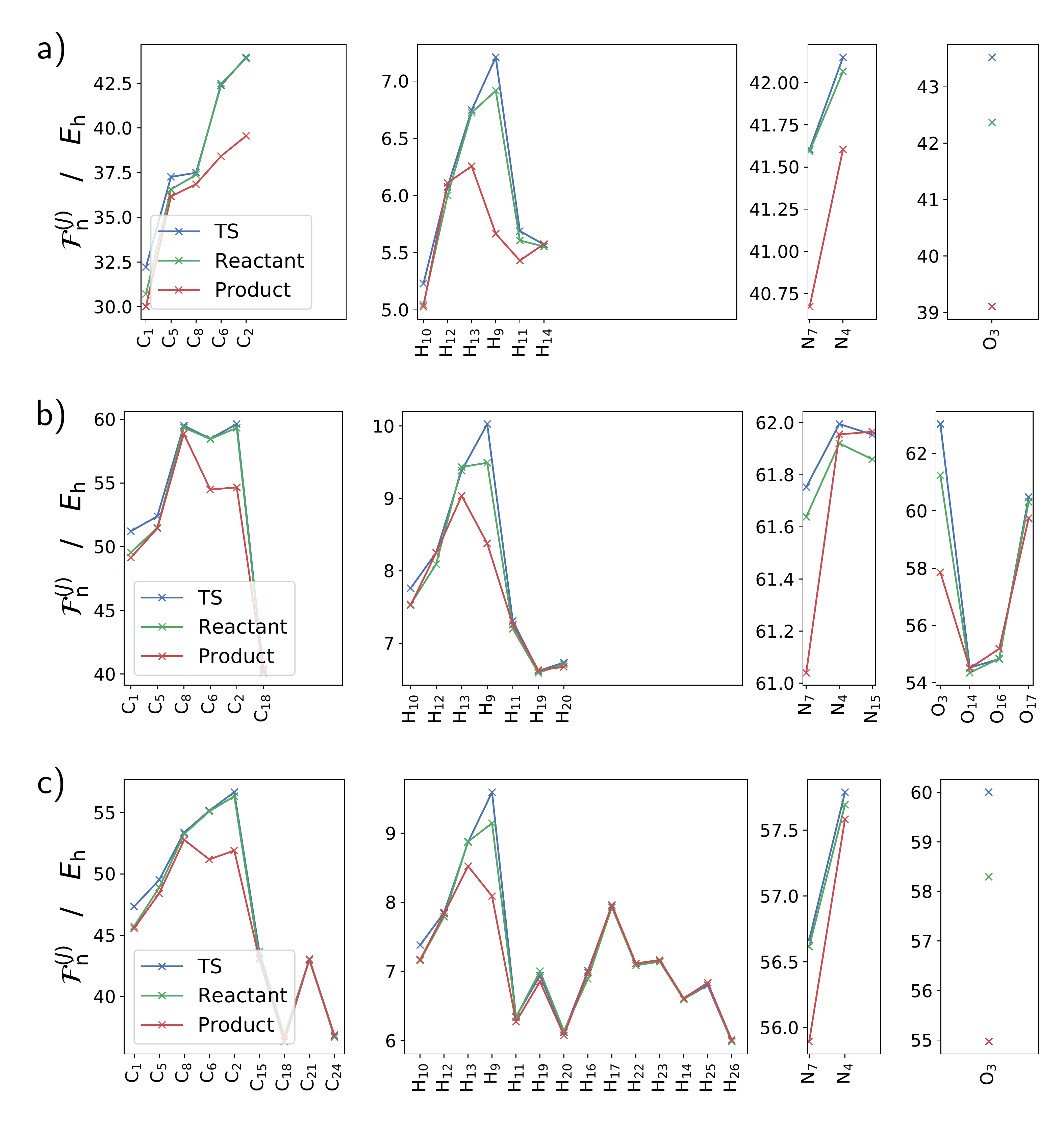}
		        \caption{For the reactions 
of Figure \ref{fig:rxn}, the nuclear repulsion features $\mathcal{F}^{(J)}_{\text{n}}$ are shown
			(a) for reaction 1 with ``generic'' H substituents,
			(b) for reaction 2 with electron-donating and -withdrawing groups \ce{NO2} and \ce{OCH3}, and
			(c) for reaction 3 two ethyl substituents
(TS denoted blue and the reactant and product as green and red, respectively).
			The plots are each split into four subplots according to the elemental composition of the reacting molecule.
		}
		\label{fig:substituents}
	\end{figure}

	To compare the effects of different substituents, we plotted the same representations as in Figure \ref{fig:relation_fnn_fne} in Figure \ref{fig:substituents}.
	In reaction 2 with the ethyl substituents, we can see that both proximal carbon atoms, \ce{C15} and \ce{C21}, are of similar magnitude, as are
	the distal carbon atoms, \ce{C18} and \ce{C21}.
	The patterns of \ce{C6} and \ce{C2} toward the product remain the same.
	Note that the relative pattern, for example, the increase from \ce{C8} to \ce{C6} in reaction 1 but the decrease in the same 
	two carbon atoms in reactions 2 and 3, should not be overinterpreted as this pattern is dependent only on the sorting and does not carry physical meaning.
	Even for reaction 3, with the electron-donating and -withdrawing groups, a similar pattern is observed.

	\begin{figure}
		\centering
		\includegraphics[width=1\textwidth]{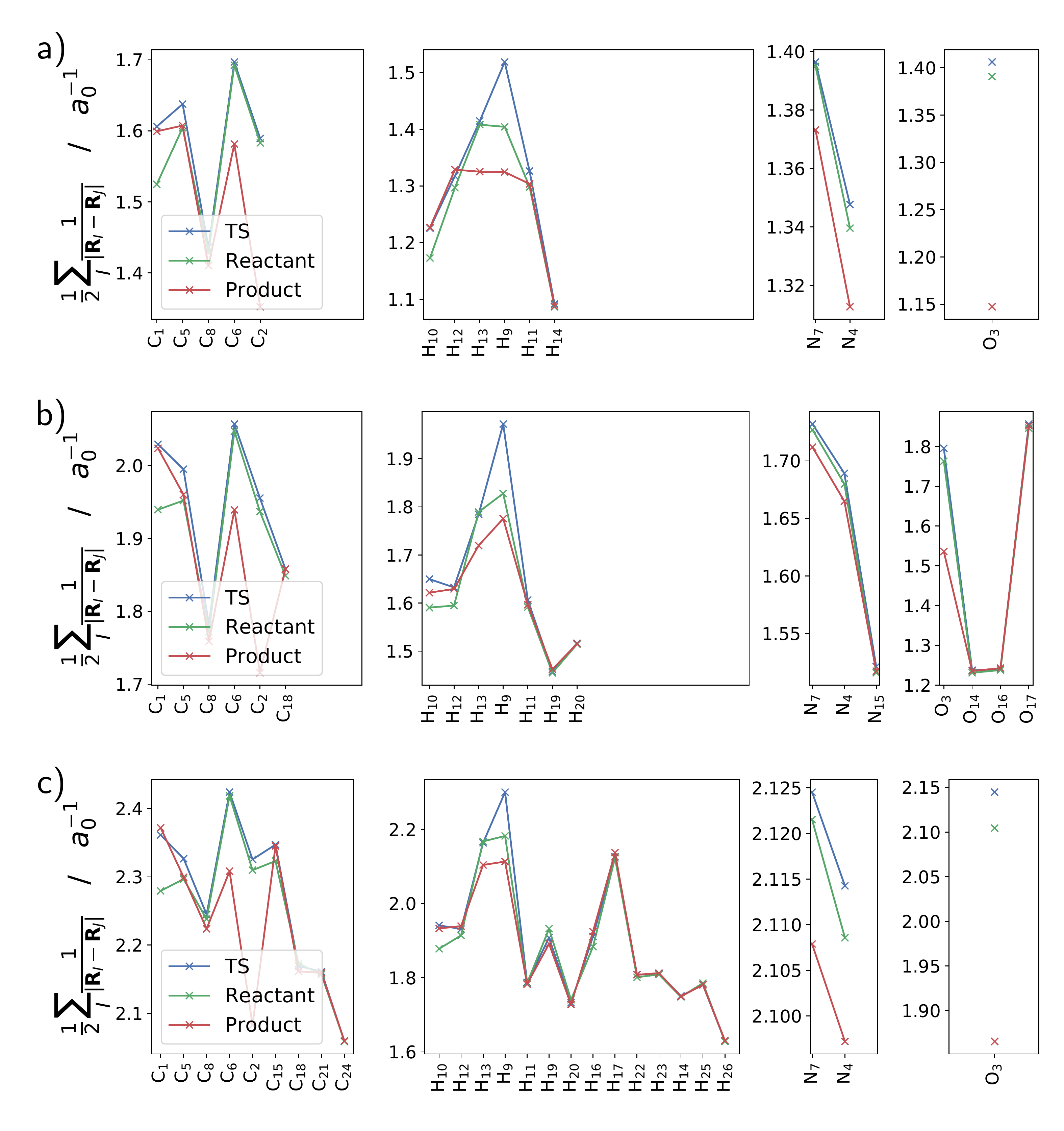}
		        \caption{For the reactions 
of Figure \ref{fig:rxn}, the inverse distance features, $\widetilde{\mathcal{F}}^{(J)}_{\text{n}}$, are shown
			(a) for reaction 1 with H substituents,
			(b) for reaction 2 with electron-donating and -withdrawing groups \ce{NO2} and \ce{OCH3}, and
			(c) for reaction 3 with two ethyl substituents
(TS denoted blue and the reactant and product as green and red, respectively). 
			The plots are each split into four subplots according to the elemental composition of the reacting molecule.
		}
		\label{fig:substituents-inv}
	\end{figure}

	\subsection{Relation of $\mathcal{F}_{\text{n}}^{(J)}$ to the CM}
	
	Consider the modified CM, $\tilde{\mathbf{C}}$, with
	diagonal elements set to 0 and the interaction terms divided by two 
	to resemble avoidance double counting:
	
	\begin{equation}
		\tilde{C}_{IJ} = 
		\begin{cases}
			0 \quad &I=J \\
			\dfrac{1}{2}\dfrac{Z_I Z_J}{|\mathbf{R}_I - \mathbf{R}_J| } \quad &I\neq J \,
		\end{cases} .
	\end{equation}
	All elements $\mathcal{F}^{(J)}_\text{n}$ can be obtained by
	multiplying the $J$-th row of this CM with 0-diagonal with a vector of one entries, $\mathbf{1}$:
	
	\begin{equation}
		\tilde{\mathbf{C}}^{(J)} \cdot
		\left[
		\begin{array}{c}
			1 \\
			1 \\
			\vdots \\
			1
		\end{array}
		\right]
		= 
		\mathcal{F}_\text{n}^{(J)} .
	\end{equation}
	
	We have $\mathcal{F}_{\text{n}}^{(J)} = \sum_{I,I\neq J} C_{IJ} = \sum_{I} \tilde{C}_{IJ}$, establishing
	a relationship to the (modified) CM $\mathbf{C}$
	($\tilde{\mathbf{C}}$).

To isolate the steric effects from the electronic effects, we consider, similarly
to Figure \ref{fig:substituents} but without the nuclear charges $Z_I$ and
$Z_J$ in the numerator, a sum of inverse distances at nucleus $J$, namely, 
$\widetilde{\mathcal{F}}^{(J)}_\text{n} \equiv \frac{1}{2} \sum_I \frac{1}{\vert
\mathbf{R}_I - \mathbf{R}_J \vert}$, in Figure \ref{fig:substituents-inv}.
This function has a larger magnitude in areas of the molecule with a higher
scaffold density that exhibit more steric effects.
For instance, as before, the function shows higher values for \ce{C6} and
\ce{C2} and a drastic change toward the product.
However, by contrast to Figure \ref{fig:substituents}, the function is characterized by a larger
value at \ce{C6} than at \ce{C2} because it strictly measures the neighbor
density, of which \ce{C6} has more (\ce{N7}, \ce{H13}, and \ce{C5} are nearby) than
\ce{C2} (only \ce{N4} and \ce{O3} are nearby).
In Figure \ref{fig:substituents}, the larger nuclear charges of 
nitrogen atoms and oxygen atoms weighted the overall sum higher.
More of such effects can be observed: another one is in reaction 2, where
the sum of inverse distances at \ce{N15} is much smaller than in Figure
\ref{fig:substituents} compared to the other nitrogen atoms, because the
neighboring oxygen atoms are not weighted according to their nuclear charges.

	\subsection{Comparison of $\mathcal{F}_{\text{n}}^{(J)}$ and CM Eigenvalues}
	
\begin{figure}
		\centering
		\includegraphics[width=1\textwidth]{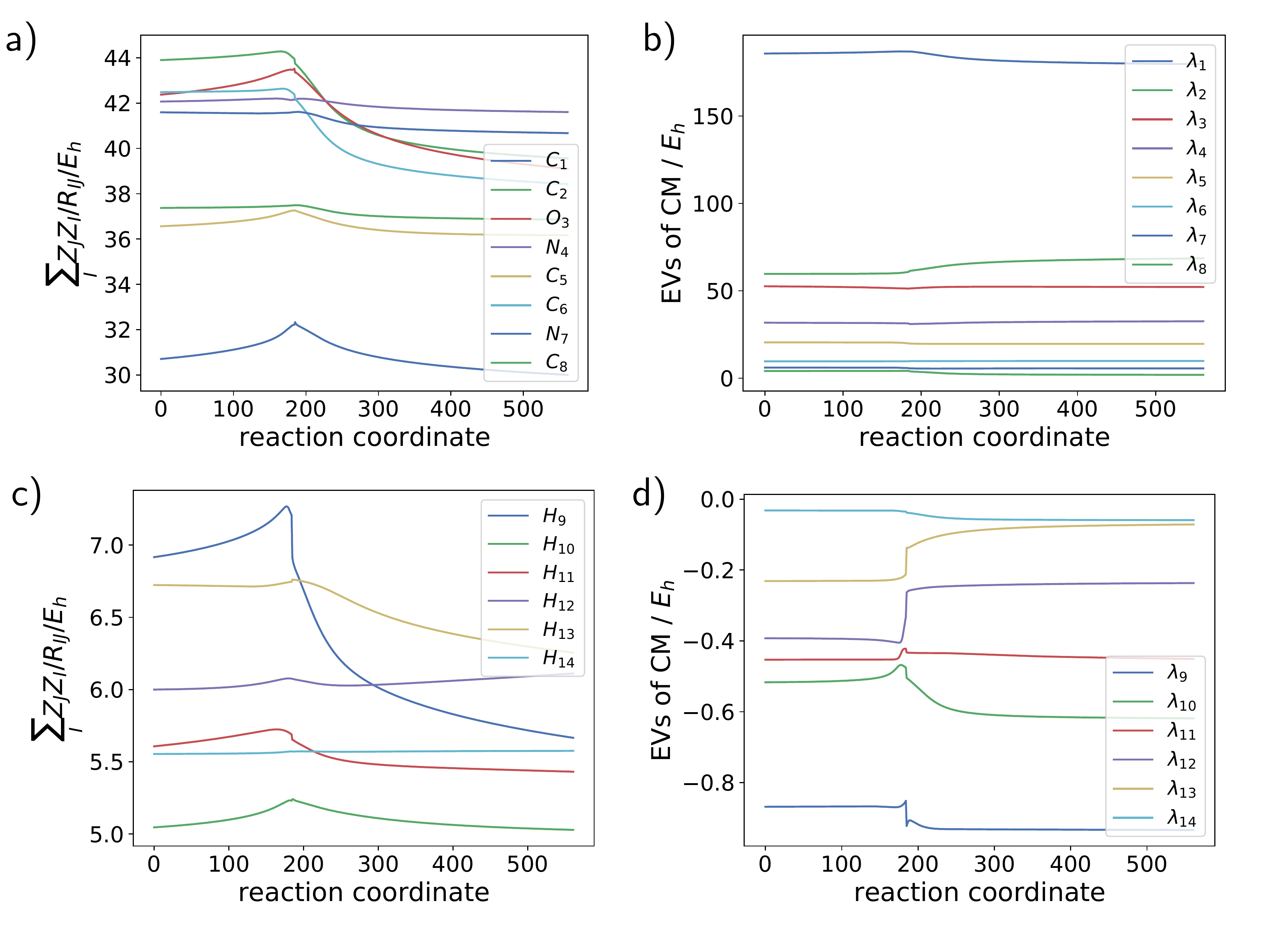}
		\caption{Traces of reaction 1 in Figure \ref{fig:rxn} over the IRC as represented
			by $\mathcal{F}_{\text{n}}^{(J)}$ (left panel) and by the eigenvalues of the CM (right panel), where each set is split in two to account for the different orders of magnitude of the data points. }
		\label{fig:traces}
	\end{figure}

	In Figure \ref{fig:traces}, the nuclear repulsion features, $\mathcal{F}_{\text{n}}^{(J)}$, are shown
	in the left panel and the eigenvalues of the CM, $\lambda_i$, in the right panel,
	all for the intrinsic reaction coordinate (IRC) of reaction 1 in Figure \ref{fig:rxn}.
	In both feature spaces, it is clearly visible what the TS structure is (step 187) and that it is most likely an early one.
	The eigenvalues of the CM are hard to interpret, however, because the values
	are not tied to a particular atom.
	By contrast, from our $\mathcal{F}_{\text{n}}^{(J)}$ descriptor, it is evident which atoms
	undergo large changes over the course of the reaction:
	the hydrogen atom that is shifted in the reaction, \ce{H9}, features the largest change
	in the trace relative to the atoms of the same element.
	This is in line with Figures \ref{fig:relation_fnn_fne} and \ref{fig:correlation_fnn_fne}, where this 
	hydrogen atom produces the largest spread.
	As before when discussing Figure \ref{fig:relation_fnn_fne}, we see that \ce{C2} and \ce{C6} behave
	similarly, as they are the bonding partners of the bond that is broken.
	The nitrogen atoms, \ce{N4} and \ce{N7}, appear relatively stable in terms of the $\mathcal{F}_{\text{n}}^{(J)}$ descriptor apart from a slight relaxation observed in all heavy atoms, due to the molecule opening up and becoming less compact.
	The carbon atom \ce{C8} with its hydrogen atom \ce{H14} also shows a rather flat trace as they
	hardly participate in the reaction.
	
	Because $\mathcal{F}_{\text{n}}^{(J)}$ yields interpretable traces and, at the same time, encodes very similar
	information to that encoded in the eigenvalues of the CM, it represents an excellent alternative,
        if not a superior type of feature,  for many applications.
	This will be especially interesting if for a given model one wants to backtrack and re-inspect a
	feature that a machine learning model paid particular attention to.
	For instance, with such information it is possible to set a threshold of change over the course
	of the reaction and filter the rest of the data set according to reactions that
	behave in a similar way.
	In each case, it is possible to go back to the descriptor and evaluate where a particular
	contribution came from.

	A drawback, however, is that $\mathcal{F}_{\text{n}}^{(J)}$ is not permutationally invariant and 
	suffers from the same issues that plague the plain CM.
	Due to this very reason, the diagonalization was introduced for the CM (see above), to achieve permutational invariance at the price
	of interpretability.
    The reasons given for diagonalizing the CM are\cite{rupp2012}: (i) the
    unique encoding; (ii) ``symmetrically equivalent atoms'' are treated the
    same; (iii) invariance with respect to permutation, translation, and
    rotation; and (iv) continuous distance. Property (i) has been discussed\cite{moussa2012,vonlilienfeld2013c} and 
    holds even for homometric molecules, (ii) is fulfilled, (iii) is a consequence
    of the ordering and not the diagonalization (there is no intrinsic ordering of
    eigenvalues as they are complex numbers and the ordering is dependent on the
    diagonalization algorithm), and (iv) is fulfilled. The sorting ensures that the
    structural properties of each molecule that are compared are of similar size.
    The caveat is, though, that these structural properties may appear for
    different chemical reasons.

        Because, in the general case, there is no way to track atoms and because this may not even be desirable (as in symmetric
        molecules certain atoms are symmetry redundant), we have to sort the list (or use any other way to introduce permutational invariance).
	Therefore, we simply sort the vector $\mathcal{F}_{\text{n}}^{(J)}$ in a descending order and ignore tracking of atoms.
        This is in analogy to the eigenvalues of the CM that are sorted the same way, but the key difference
        remains: the entries are still atom specific and do not mix information from all atoms (as the diagonalization
        of the CM does).
	
    Similar to our CL, in the sorted CM,\cite{hansen2015} each element of the matrix is part of the feature, whereas the rows (or columns of this symmetric matrix) are ordered according to their norm. This is equivalent to our approach apart from the fact that there is no summation over the rows (or columns, respectively), leading to a high dimensional feature, which is quadratic in the number of atoms. Moreover, it has been noted that slight variations in atomic coordinates may cause abrupt changes in the CM ordering, thereby impeding the learning of structural similarities \cite{hansen2013}. 
	
	Schrier found that the eigenvalues of the CM will not be able to distinguish larger molecules \cite{schrier2020}.
	We do not expect that this is different for the sorted $\mathcal{F}_{\text{n}}^{(J)}$.
	Furthermore, for macromolecules of more than 10,000 atoms, not only is the
    global description not granular enough, but also 
    the diagonalization needed for the eigenvalues becomes a true computational
    bottleneck, as it scales with $\mathcal{O}(M^3)$, while subsequent sorting
    is affected by a negligible cost of order $\mathcal{O}(M \log M)$, where
    $M$ is the number of nuclei.

	\subsection{Descriptor Extensivity}
	
	One long-standing issue in descriptor research is the desired transferability between molecules of different sizes. 
	Hence, an intensive descriptor is sought for, that is, one that is independent of molecular size.
	However, this desire contrasts the inherent extensivity of structures.
	One approach is to find sub-information that is intensive.
	
	The size (number of elements) of the nuclear repulsion features, $\mathcal{F}_{\text{n}}^{(J)}$,
	grows with system size, $M$.
	For the specific case of an elementary reaction step, only a very localized part of the system will react, while 
	most of the system remains largely unchanged (i.e., internal coordinates of observer atoms change only little), 
        as illustrated by 
	the observer hydrogen atom \ce{H14} in Figure \ref{fig:traces}.
	A simple solution to this problem is to truncate the feature and to consider only a subset of atoms that vary
	more than a given threshold and can be considered a relevant subsystem for the process under consideration.
        More elaborate measures may consider features evaluated in an embedding framework.

	However, the issue remains that even if two feature vectors are truncated to the same size
	to contain only the physically relevant part, they cannot be compared in a direct manner.
	Only if the set of atoms to which the vector has been truncated is the same for every
	system, a direct comparison is possible.
	The dilemma is that, as soon as convolutions are introduced to achieve permutational
	invariance, the comparability fades away.

\section{Smooth Overlap of Atomic Positions and of Electron Densities}

	We are advised to compare the results obtained so far to one of the most successful representations for molecular similarity: the SOAP kernel\cite{bartok2013}.
	In this section, we review the key derivation steps of the SOAP kernel as we need them later to
        formulate and evaluate our electron-density-based descriptor. In our derivation, we follow Ref. \citenum{bartok2013} but extend it at key places to highlight important steps of the derivation that are not explicit in the original paper and that become important for our electron-density-based descriptor, for which the derivation must be made more transparent.
For this electron-density-based descriptor, we will reinterpret the fuzzy
atomic positions of SOAP and generalize them to the actual electron density.
Accordingly, we call the resulting descriptor SOED.
We emphasize again that the derivation reviewed for SOAP in the next section is necessary as it will turn out to be the
key evaluation strategy within our setting to evaluate SOED. 

\subsection{SOAP Kernel}
	\label{ssec:repr}

SOAP represents a measure for molecular similarity without making any direct connection to the associated electronic energies (by contrast to the CM). 
	A molecule is put with its center of mass at the origin of the coordinate system. We place on each of its atoms $I$ at position $\mathbf{R}_I$ an unnormalized Gaussian function, $g_{\alpha_I}$, with parameter $\alpha = 1/(2{\sigma_I}^2)$, where $\sigma_I$ is the variance of the Gaussian function at atom $I$, to obtain a superposition of fuzzy atomic positions
	
	\begin{equation}
		\rho_{\text{mol}}(\mathbf{r};\{\mathbf{R}_I\}, \{\alpha_I\})
		= N_\mathrm{mol}
		\sum_{I=1}^M
		g_{\alpha_I}(\left| \mathbf{r} - \mathbf{R}_I \right|) 
		\label{eqn:mol-dens}
	\end{equation}
	
	with the unnormalized Gaussian
	
	\begin{equation}
		g_{\alpha_I}(\left| \mathbf{r} - \mathbf{R}_I \right|) 
		= \exp
		\left(
		-\alpha_I\left|\mathbf{r}-\mathbf{R}_I\right|^{2}
		\right)
		\label{eqn:g}
	\end{equation}
	
	and normalization constant $N_\mathrm{mol}$ to ensure that $\langle \rho_{\text{mol}} \vert \rho_{\text{mol}} \rangle = 1$. The variable $\mathbf{r}$ is the variable of the field.
	Note that in the original publication\cite{bartok2013}, the normalization constant $N_\mathrm{mol}$ is omitted because later (see below) the kernel will be normalized.
	
	In the original paper\cite{bartok2013}, this superposition has been called ``atomic neighbor density'' that represents the ``atomic environment'', which we do not adopt here as the superposition in Eq. (\ref{eqn:mol-dens}), primarily, does not put an emphasis on some local atomic structure so that other atoms become neighbors or an environment. Instead, it refers to the molecular structure as a whole.
	Hence, we may refer to it as a ``molecular-scaffold density''. This is also advantageous to conceptually emphasize its
        relation to our SOED descriptor to be introduced later.
	The comparison of two molecular-scaffold densities then requires the
    definition of an overlap measure (see below), which is the origin of the term ``smooth overlap of atomic positions'', 
which we understand as an overlap of molecular scaffolds represented by fuzzy atomic positions.
	
	Usually, the $\alpha_I$ in Eq. (\ref{eqn:mol-dens}) is taken to be the same for all atoms, $\alpha_I\rightarrow\alpha$.
	In some applications, $\alpha$ was optimized as a hyperparameter\cite{de2016,raimbault2019} or simply fixed to some value\cite{deringer2018,ferre2017}.
	Using a different $\alpha$ as a hyperparameter for each element
    type\cite{engel2019} resulted in the same value for all types during the
    training procedure. 
	
\begin{figure}[h]
		\centering
		\includegraphics[width=0.6\textwidth]{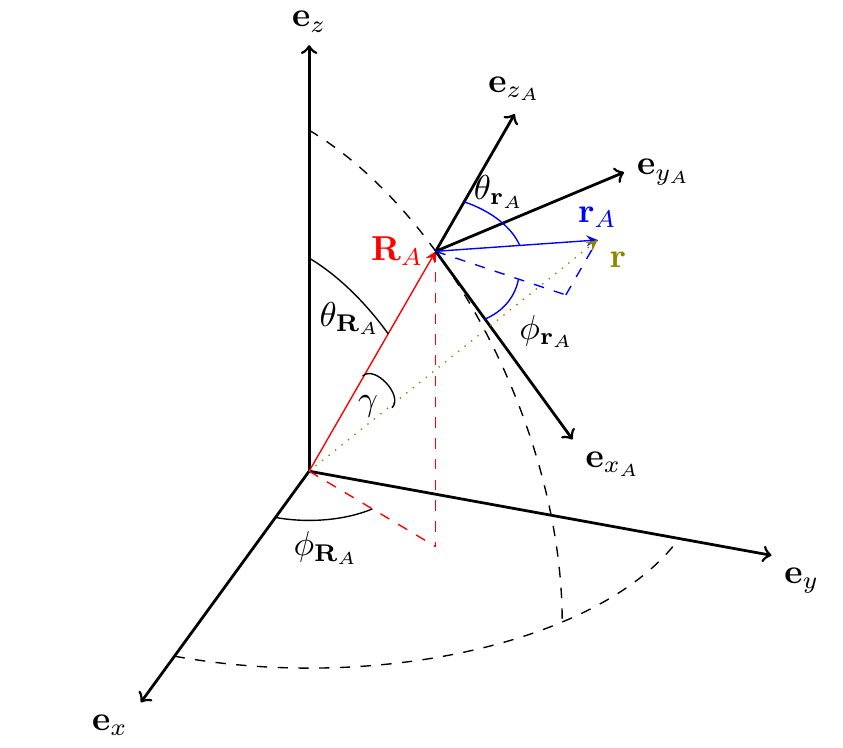}
		\caption{In the global coordinate system, $\mathbf{e}_x$,
        $\mathbf{e}_y$, and $\mathbf{e}_z$, the angles $\theta$ and $\phi$
    denote the polar and azimuthal angles of the field variable $\mathbf{r}
\equiv \mathbf{r}(\theta,\phi)$, labeled in green. Nucleus $A$ is at position
$\mathbf{R}_A\equiv\mathbf{R}_A(\theta,\phi)$ with polar and azimuthal angles $\theta_{\mathbf{R}_A}$ and $\phi_{\mathbf{R}_A}$. In the coordinate system that is centered on nucleus $A$, namely $\mathbf{e}_{x_A}$, $\mathbf{e}_{y_A}$, and $\mathbf{e}_{z_A}$, the distance to the field variable is $\mathbf{r}_A = \mathbf{r}-\mathbf{R}_A$ with corresponding polar angles $\theta_{\mathbf{r}_A}$ and $\phi_{\mathbf{r}_A}$. The angle between $\mathbf{R}_A$ and $\mathbf{r}$ is denoted $\gamma$.}
		\label{fig:multicenter}
	\end{figure}

	Notably, the sum in Eq. (\ref{eqn:mol-dens}) is permutationally invariant, a property that is important for machine learning: exchanging the terms in the sum does not change the total density. 
    In order to determine the best match of the scaffold densities of two
    molecules for their comparison regarding the assessment of molecular
    similarity, it will be necessary to rotate one scaffold density in a
    three-dimensional space with respect to the other. To be able to rotate the
    sum of Gaussians in Eq.  (\ref{eqn:mol-dens}), the equation must be
    expanded into functions dependent on the global polar angles, $\theta$ and
    $\phi$ (see Figure \ref{fig:multicenter}).\cite{bartok2013} In the definition above, they are dependent on $\mathbf{r}_A$, which refers to the frame of reference that is centered on nucleus $A$ and that is not easily rotatable from a global point of view.  The functions sought for straightforward rotation of the whole molecular scaffold field\cite{bartok2013} are all to be located at the same  single center, which can be the center of mass coordinates of the molecule. 
	
	When we multiply out the exponent in Eq. (\ref{eqn:mol-dens}), we obtain
    $-\alpha (\mathbf{r}^2 - 2 \, \mathbf{r} \cdot \mathbf{R}_I + \mathbf{R}_I^2)$, that is, two squared terms and a cross term. The cross term is then subjected to a Rayleigh expansion of a plane wave in terms of spherical waves,\cite{bartok2013}
	\begin{equation}
		\exp \left(\mathbf{r} \cdot \mathbf{R}_{I}\right)=\sum_{l=0}^{\infty}(2 l+1) i_{l}\left(r R_{I}\right) P_{l}(\cos \gamma)
		\label{eqn:rayleigh}
	\end{equation}
	where $i_l$ is the modified spherical Bessel function of the first kind of degree $l$, $P_l$ is the $l$th Legendre polynomial, and $\gamma = \angle(\mathbf{r},\mathbf{R}_I)$, measured from the origin of the coordinate system (see Figure \ref{fig:multicenter}). 
	The modified spherical Bessel function of the first kind, $i_{n}(x) \equiv \sqrt{\frac{\pi}{2 x}} I_{n+1 / 2}(x)$, is one radial solution to the Helmholtz equation in spherical coordinates and related to the modified Bessel function of the first kind,  $I_{n}(x) \equiv i^{-n} J_{n}(i x)$, where $J_n$ in turn is the Bessel function of the first kind, which is a solution to Bessel's differential equation, other solutions being the Bessel function of the second kind and Hankel functions. 
	Hence, the molecular scaffold density in Eq. (\ref{eqn:mol-dens}) can be expressed as
	\begin{equation}
		\begin{aligned}
			\rho_\text{mol}( \mathbf{r}; \{ \mathbf{R}_I \}, \alpha) 
			&= \sum_I^M \exp \left(-\alpha\left|\mathbf{r}-\mathbf{R}_{I}\right|^2\right)\\
			&= \sum_I^M \exp \left(-\alpha (\mathbf{r}^2 - 2\mathbf{r R}_I + \mathbf{R}_I^2)\right)\\
			&= \sum_I^M \exp \left(-\alpha\left(r^2+R_{I}^2\right)\right) \sum_{l=0}^{\infty}(2 l+1) i_{l}\left(2 \alpha r R_{I}\right) P_{l}(\cos \gamma) \ ,
		\end{aligned}
	\end{equation}
	where the normalization constant $N_\text{mol}$ was omitted as explained above. By virtue of the spherical harmonics addition theorem,
	\begin{equation}
		P_{l}(\cos \gamma) 
		=\frac{4 \pi}{2 l+1} 
		\sum_{m=-l}^{l} 
		{Y}_{lm}^{\ast}(\theta_{\mathbf{R}_I},\phi_{\mathbf{R}_I})
		Y_{lm}(\theta,\phi) 
		\label{eqn:shat}
	\end{equation}
	with angles $\theta$ and $\phi$ as the polar coordinates of the field variable $\mathbf{r}$ measured from the origin and the angles $\theta_{\mathbf{R}_I}$ and $\phi_{\mathbf{R}_I}$ being the polar coordinates of $\mathbf{R}_I$, 
notably also measured from the new common origin. These steps accomplish a single-center expansion necessary for the subsequent rotation in search of the best matches of two molecular scaffold densities. We now obtain a compact expression of expansion coefficients and spherical harmonics, where the term $(2l+1)$ cancels out
	\begin{equation}
		\begin{aligned}
			\rho_\text{mol}(\mathbf{r}; \{\mathbf{R}_I \}, \alpha)
			&= 4 \pi 
			\sum^M_{I} 
			\exp (-\alpha(r^2 + R_{I}^2))
			\sum_{l=0}^{\infty} 
			\sum_{m=-l}^{+l} 
			i_{l}\left(2 \alpha r R_{I}\right) 
			Y_{l m}^{*}(\theta_{\mathbf{R}_I},\phi_{\mathbf{R}_I})
			Y_{l m}(\theta,\phi) \\
			&= \sum^M_{I} \sum_{l m} c_{l m}^{I}(r; \mathbf{R}_I, \alpha) Y_{l m}(\theta,\phi)
			\label{eqn:bessel-expansion}
		\end{aligned}
	\end{equation}
	with
	\begin{equation}
		c_{l m}^{I}(r;\mathbf{R}_I, \alpha)
        \equiv 4 \pi \exp \left(-\alpha\left(r^2+R_{I}^2\right)\right)
		i_{l}\left(2 \alpha r R_{I}\right)
		Y_{l m}^{*}(\theta_{\mathbf{R}_I}, \phi_{\mathbf{R}_I} ) \ .
		\label{eqn:expansion-coef}
	\end{equation}
	
    where $\sum_{lm}$ denotes $\sum_l^\infty \sum_{m=-l}^{l}$.
	Note that the angular information of the nuclear position is stored in the spherical harmonics 
$Y_{l m}^{*}(\theta_{\mathbf{R}_I}, \phi_{\mathbf{R}_I} )$. It is exactly these functions that we will need to
rotate against each other to generate best overlaps over all possible rotations
of the two molecular scaffold densities to be compared. This is due to the fact that the 
density field is defined in terms of Gaussian functions located at the positions of the nuclei, which therefore need to be rotated.
We emphasize this point because it is important for our SOED derivation below and because it appears somewhat obscured in
the original paper\cite{bartok2013}. 
The overlap between these two densities, $\rho_\text{mol}(\mathbf{r};\mathbf{R}_I, \alpha)$ and $\rho_\text{mol}^\prime(\mathbf{r};\mathbf{R}_{I^\prime}, \alpha)$, is defined as the inner product of their densities, 
	
	\begin{equation}
		S(\rho_\text{mol}, \rho_\text{mol}')
		= \langle \rho_\text{mol} \mid \rho_\text{mol}' \rangle 
		= \int_{\R^3} \mathrm{d}\mathbf{r} \rho_\text{mol} \rho_\text{mol}' \ ,
		\label{eq:overlap-mol}
	\end{equation}
	
where we omitted the parameter dependence for the sake of clarity
	and $\mathrm{d}\mathbf{r} = \mathrm{d}x \, \mathrm{d}y \, \mathrm{d}z$.
	The center of mass of both molecules matches the origin of the coordinate system, leaving rotational freedom around three Euler angles, $\alpha,\beta,\gamma$. 
Hence, in search for optimal overlaps of the two molecular-scaffold densities
we need to integrate over all possible rotations.\cite{bartok2013}
	This procedure automatically guarantees rotational invariance.
	
	A rotation from the rotation group, $\mathrm{SO}(3)$, can be written as a matrix,
	\begin{equation}
		\hat{R} \equiv \hat{R}(\alpha, \beta, \gamma) \ ,
	\end{equation}
	where $\alpha, \beta, \gamma$ are the Euler angles. 
	This rotation operator needs to be integrated over SO(3) when acting on a density; that is,
all possible rotations of the density must be considered. 
The volume element on SO(3) 
integrating over all possible Euler angles yields the measure
	
	\begin{equation}
		\mathrm{d}\Omega = \frac{1}{8 \pi^{2}} \, \sin\beta \, \mathrm{d}\alpha \, \mathrm{d}\beta \, \mathrm{d}\gamma
	\end{equation}
	
	with $\alpha = [0, 2 \pi)$, $\beta = [0, \pi)$, and $\gamma = [0, 2 \pi)$.

	We now need to rotate the spherical harmonic $Y_{l m}(\theta_{\mathbf{R}_I}, \phi_{\mathbf{R}_I})$ in the expansion in Eq. (\ref{eqn:bessel-expansion}).
	An arbitrary rotation operator, $\hat{R}$, operating on a spherical harmonic yields a linear combination of new spherical harmonics with a different magnetic quantum number $m$ and the elements of a Wigner D-matrix as expansion coefficients. 
	With this identity, we will rotate the angular information of the nuclear positions (because the radial part is separated) according to
	
	\begin{equation}
		\hat{R} Y_{l m}(\theta_{\mathbf{R}_I}, \phi_{\mathbf{R}_I})
		=\sum_{m'=-l}^{l} 
		D_{m m'}^{l}(\hat{R}) 
		Y_{l m'}(\theta_{\mathbf{R}_I}, \phi_{\mathbf{R}_I}) \ .
		\label{eqn:wigner-rotation}
	\end{equation}
	
	The elements of the Wigner matrices are given by 
	
	\begin{equation}
		D_{m m'}^{l}(\hat{R})
		=\left\langle Y_{l m}(\theta_{\mathbf{R}_I}, \phi_{\mathbf{R}_I})
		\left| \, \hat{R} \, \right| 
		Y_{l m'}(\theta_{\mathbf{R}_I}, \phi_{\mathbf{R}_I})\right\rangle \ .
	\end{equation}
	This is convenient as the rotation happens in a very compact manner without
    notational overhead usually involved in rotations. Furthermore, we will later see how multiple Wigner D-matrices will cancel each other out.
	
	When considering the overlap of two molecular-scaffold densities with one of them being rotated,
	
	\begin{equation}
		\begin{aligned}
			S\left(\rho_\text{mol}, \hat{R}\rho_\text{mol}'\right)
			= \langle \rho_\text{mol} \mid \hat{R} \mid \rho_\text{mol}' \rangle 
			= \int_{\mathbb{R}^3} d\mathbf{r} \rho_\text{mol} \hat{R} \rho_\text{mol}'
			\label{eqn:overlap}
		\end{aligned}
	\end{equation}

	we will have to integrate over all Euler angles $\alpha,\beta,\gamma$, which will yield
	the rotationally invariant kernel,
	
	\begin{equation}
		\begin{aligned}
			k\left(\rho_\text{mol}, \rho_\text{mol}'; n\right)
			=\int_{\alpha,\beta,\gamma} 
			\left| 
			\int_{\mathbb{R}^3} d\mathbf{r} \rho_\text{mol} \hat{R} \rho_\text{mol}'
			\right|^{n}
			\frac{1}{8 \pi^{2}} \sin\beta \,
			\mathrm{d}\alpha \, \mathrm{d} \beta \, \mathrm{d}\gamma \\
			\label{eqn:kernel1}
		\end{aligned}
	\end{equation}
	
	where we follow the original paper\cite{bartok2013} and artificially introduced the power $n$ as a parameter to be considered later; for now, we set it equal to 1. As we will see in the following, this integral over Euler angles can be evaluated analytically with help of the Wigner D-matrices and will not require us to deal with the explicit rotation algebra as shown in Eq. (\ref{eqn:wigner-rotation}).
	
	We substitute the expanded density of Eq. (\ref{eqn:bessel-expansion}) into the overlap Eq. (\ref{eqn:overlap}) to obtain the overlap of the two molecular-scaffold densities, where we exploit the rotation identity of Eq. (\ref{eqn:wigner-rotation}), 
	
		\begin{eqnarray}
			S\left(\rho_\text{mol}, \hat{R} \rho_\text{mol}'\right)
			&=& \int d \mathbf{r} \rho_\text{mol} \hat{R}\rho_\text{mol}' \nonumber\\
			&=& \int d \mathbf{r} 
			\rho_\text{mol}
			\Big[
			\sum_{I'}^{M'} 
			\sum_{l' m'} 
			4 \pi 
			\exp [-\alpha_{I'}(r^2+R_{I'}^2)]
			i_{l'}(2 \alpha_{I'} r R_{I'})
			\hat{R}
			Y_{l' m'}^{*}(\Omega_{\mathbf{R}_{I'}})
			Y_{l' m'}(\Omega) 
			\Big] \nonumber\\
			&=& \int d \mathbf{r} 
			\rho_\text{mol}
			\Big[
			\sum_{I'}^{M'} 
			\sum_{l' m'} 
			4 \pi 
			\exp [-\alpha_{I'}(r^2+R_{I'}^2)]
			i_{l'}(2 \alpha_{I'} r R_{I'}) \nonumber\\
			&&\times\sum_{m''=-l'}^{l'}
			D_{m' m''}^{l'}(\hat{R}) 
			Y_{l' m''}^{*}(\Omega_{\mathbf{R}_{I'}})
			Y_{l' m'}(\Omega) 
			\Big] 
		\label{eqn:integral-1}
		\end{eqnarray}
	
	where a new coefficient, analogously to Eq. (\ref{eqn:expansion-coef}) with different indices,
	
	\begin{equation}
		c_{l' m''}^{I'} =
		4 \pi 
		\exp [-\alpha_{I'}(r^2+R_{I'}^2)]
		i_{l'}(2 \alpha_{I'} r R_{I'})
		Y_{l' m''}^{*}(\Omega_{\mathbf{R}_{I'}})
	\end{equation}
	
	can be employed to collect many of the terms. Substituting this coefficient and separating the integral into a radial and a spherical part, we obtain
	
		\begin{eqnarray}
			S\left(\rho_\text{mol}, \hat{R} \rho_\text{mol}'\right)
			&=&  \sum_{I, I'}^{M, M'}
			\sum_{l, m \atop l', m'} 
			\sum_{m''=-l'}^{l'}
			D_{m'm''}^{l'}(\hat{R})
			\underbrace{
				\int_{r=0}^{\infty} \mathrm{d}r
				r^2 
				c_{l m}^{I *} 
				c_{l' m''}^{I'} 
			}
			_{\mathrm{radial}}\nonumber\\
			&&\times \underbrace{
				\int_{\theta=0}^{2\pi}\int_{\phi=0}^{\pi}  
				\sin \phi \, \mathrm{d}\phi \, \mathrm{d}\theta \, 
				Y_{l m}^{*}(\Omega) 
				Y_{l' m'}(\Omega)}
			_{\mathrm{spherical}}
		\label{eqn:integral-2}
		\end{eqnarray}

	where we abbreviated the angular information 
according to $\Omega = (\theta,\phi)$ and $\Omega_{\mathbf{R}_{I}} = (\theta_{\mathbf{R}_{I}}, \phi_{\mathbf{R}_{I}})$. 
Recall that we also chose to have a general $\alpha_I$ for the unprimed and $\alpha_{I'}$ for the primed molecular scaffolds.

	Now, the integral in Eq. (\ref{eqn:integral-2}) must be evaluated. Because the spherical harmonics are orthonormal by definition, the {spherical part} in Eq. (\ref{eqn:integral-2}) evaluates to 
	
	\begin{equation}
		\int_{\theta=0}^{2\pi}\int_{\phi=0}^{\pi}  
		\sin \phi \, \mathrm{d}\phi \, \mathrm{d}\theta \, 
		Y_{l m}^{*}(\Omega) 
		Y_{l' m'}(\Omega)
		= \delta_{ll'} \delta_{m m'} \ .
	\end{equation}
	
This simplifies the last line in Eq. (\ref{eqn:integral-2}) to the following, 
	
	\begin{equation}
		S\left(\rho_\text{mol}, \hat{R} \rho_\text{mol}'\right) =
		\sum_{I, I'} 
		\sum_{l, m} 
		\sum_{m''=-l}^{l}
		D_{mm''}^{l}(\hat{R})
		\int_{r=0}^\infty \mathrm{d} r r^2 c_{l m}^{I *} c_{l m''}^{I'}\ . 
		\label{eqn:integral-simple}
	\end{equation}
setting $m \leftarrow m'$ and $l \leftarrow l'$. 
	
	The {radial part} of the integral in Eq. (\ref{eqn:integral-2}) and Eq. (\ref{eqn:integral-simple}) is non-trivial. We first expand it to move the quantities independent of the position before the integral,
	
	\begin{equation}
		\begin{aligned}
			\int_{r=0}^\infty 
				\mathrm{d}r 
				r^2 
				c_{l m}^{I*} 
				c_{l m''}^{I'}
			&=
			(4 \pi)^2
			\exp(-(\alpha_{I'}R_{I'}^2 + \alpha_{I}R_{I}^2))
			Y_{l m}\left(\Omega_{\mathbf{R}_{I}}\right)
			Y_{l^{\prime} m^{\prime \prime}}^{*}\left(\Omega_{\mathbf{R}_{I^{\prime}}}\right) \\
			& \times
			\int_{r=0}^\infty 
				\mathrm{d}r 
				r^2 
				\exp(-(\alpha_{I'} + \alpha_{I}) r^2 ) 
				i_l   (2 \alpha_I    r  R_I)
				i_{l'}(2 \alpha_{I'} r  R_{I'})
		\end{aligned}
		\label{eqn:radial-part}
	\end{equation}

where the integral is solved as

	\begin{equation}
	\begin{aligned}
		\int_{r=0}^\infty 
			&
			\mathrm{d}r 
			r^2 
			\exp(-(\alpha_{I'} + \alpha_{I}) r^2 ) 
			i_l   (2 \alpha_I    r  R_I)
			i_{l'}(2 \alpha_{I'} r  R_{I'}) \\
			&= 
			\frac{1}{4}
			\left(
				\frac{\pi}
				{(\alpha_I+\alpha_{I'})^{3}}
			\right)^{1 / 2} 
			i_{l}\left(
				2 \frac{\alpha_{I} \alpha_{I'}}{\alpha_{I}+\alpha_{I'}} 
				R_{I} R_{I'}
			\right) 
			\exp \left(
				\frac{\alpha_{I}^{2} R_{I}^{2} + \alpha_{I'}^{2} R_{I'}^{2}}{\alpha_{I}+\alpha_{I'}}
			\right)
			\label{eqn:weber-int}
		\end{aligned}
	\end{equation}
	according to Weber.\cite{watson1995,weber1868}
	
	The prefactor in Eq. (\ref{eqn:radial-part}) can be combined with the result from the integral Eq. (\ref{eqn:weber-int}).
    If we had considered the two normalization constants $N_\mathrm{mol}=(\frac{2 \alpha}{\pi})^{3/4}$ that we omitted from Eq. (\ref{eqn:mol-dens}), we would have obtained the same overlap as Kaufmann and Baumeister,\cite{kaufmann1989}
	
	\begin{equation}
			S\left(\rho_\text{mol}, \hat{R} \rho_\text{mol}'\right)
			=
			\sum_{I, I'} 
			\sum_{l, m, m''} 
			D_{mm''}^{l}(\hat{R})
			\tilde{I}_{mm''}^l(\alpha_I, \alpha_{I'}, \mathbf{R}_{I}, \mathbf{R}_{I'})
	\label{eqn:soap-overlap}
	\end{equation}
	
	with 
	
	\begin{equation}
	\begin{aligned}
		\tilde{I}_{mm''}^l(\alpha_I, \alpha_{I'}, \mathbf{R}_{I}, \mathbf{R}_{I'})
		&=
		4 \pi
		\left(
		2 \frac{
			(\alpha_I \alpha_{I'})^{1 / 2}
		}{
			\alpha_I+\alpha_{I'}}
		\right)^{3 / 2} 
		\exp \left(
		-\frac{\alpha_I \alpha_{I'}}{\alpha_I+\alpha_{I'}}(R_{I}^{2}+R_{I'}^{2})
		\right) \\
		& \times
		i_{l}\left(
		2 \frac{\alpha_I \alpha_{I'}}{\alpha_I+\alpha_{I'}} 
		R_{I} R_{I'}
		\right) 
		Y_{l m}(\Omega_{\mathbf{R}_{I}}) 
		Y_{l' m''}^{*}(\Omega_{\mathbf{R}_{I'}}) \ .
		\end{aligned}
	\end{equation}

	For $\alpha \leftarrow \alpha_I$ and $\alpha \leftarrow \alpha_{I'}$, that
    is, $\tilde{I}(\alpha, \alpha, \mathbf{R}_{I}, \mathbf{R}_{I'})$, we
    recover the result obtained in the original SOAP paper,\cite{bartok2013} which was there denoted as $\tilde{I}_{m m^{\prime\prime}}^{l}\left(\alpha, \mathbf{R}_{I}, \mathbf{R}_{I'}\right)$. The prefactor on the original paper, $\sqrt{\frac{2 \pi^{5}}{\alpha^{3}}}$, as noted in the erratum\cite{bartok2017b}, 
    originates from the missing normalization constants, $N_\mathrm{mol}$. If
they were included, that new prefactor would not have been necessary, as we can
easily verify by $\sqrt{\frac{2 \pi^{5}}{\alpha^{3}}}  N_\mathrm{mol}^2 = 4\pi$,
where $4 \pi$ is
their original prefactor. As already noted in the erratum\cite{bartok2017b},
this does not produce an error, as it gets cancelled at the normalization step.
		
	In the original paper\cite{bartok2013}, the authors defined a term for the sum over all pairwise interactions of the integral 
	
	\begin{equation}
		I_{m m''}^{l} 
		\equiv 
		\sum_{I, I'} 
		\tilde{I}_{m m''}^{l}(\alpha, \mathbf{R}_{I}, \mathbf{R}_{I'}) .
	\end{equation}
	
	to obtain a slightly more succinct form
	
	\begin{equation}
		S\left(\rho_\text{mol}, \hat{R} \rho_\text{mol}'\right) 
		=\sum_{l, m, m''} I_{m m''}^{l} D_{m m''}^{l}(\hat{R})
	\end{equation}
	
	which we will not adapt here as it obfuscates the sums that nicely emphasize the series expansion over the indices $l$, $m$, and $m'$ as well as the pairwise nuclear interaction
over the indices $I$ and $I'$.

	We recall from Eq. (\ref{eqn:kernel1}) that the integral of the overlap defines the kernel.
	For $n=2$, the rotationally invariant kernel is 
	
	\begin{equation}
		\begin{aligned}
			k\left(\rho_\text{mol}, \rho_\text{mol}';2\right) 
			&=\int \left| S\left(\rho_\text{mol}, \hat{R} \rho_\text{mol}'\right) \right| ^2 \mathrm{d}\Omega \\
			&=\int \mathrm{d} \Omega \, 
			S\left(\rho_\text{mol}, \hat{R} \rho_\text{mol}'\right)^\ast 
			S\left(\rho_\text{mol}, \hat{R} \rho_\text{mol}'\right) \\
			&=\sum_{I, I'}
			\sum_{l, m, m' \atop \lambda, \mu, \mu'}
			\tilde{I}_{m m'}^{l*} 
			\tilde{I}_{\mu \mu'}^{\lambda}
			\underbrace{
			\int \mathrm{d}\Omega \,
				D_{m m'}^{l}(\hat{R})^{*}
				D_{\mu \mu'}^{\lambda}(\hat{R})
			}_{\text{Wigner's orth. relation}}\\
			&=\sum_{I, I'}\sum_{l, m, m'}\dfrac{1}{2l+1} \tilde{I}_{m m'}^{l*} \tilde{I}_{m m'}^{l}
		\end{aligned}
		\label{eqn:final-ints}
	\end{equation}
	
	where we used new indices to avoid the doubly primed $m''$ and where by virtue of Wigner's orthogonality relation, 
        we have
	
	\begin{eqnarray}
		\int_{0}^{2 \pi} \mathrm{d} \alpha 
		\int_{0}^{\pi} \mathrm{d} \beta \sin \beta 
		\int_{0}^{2 \pi} 
                & \mathrm{d} \gamma  &
		\frac{1}{8 \pi^2}
		D_{m' k'}^{j'}(\hat{R}(\alpha, \beta, \gamma))^{*} 
		D_{m k}^{j}(\hat{R}(\alpha, \beta, \gamma)) \nonumber\\
		&=&\frac{1}{2 j+1} \delta_{m' m} \delta_{k' k} \delta_{j' j}
	\end{eqnarray}
	
We see that the Greek indices collapse with the Latin ones, $m \leftarrow \mu$, $m' \leftarrow \mu'$, and $l \leftarrow \lambda$.
	
    The kernel is eventually defined with a power $\zeta$, which is also a
    hyperparameter as it steers the sensitivity to the kernel changing the atomic positions, and we set it to unity to not obfuscate the results further. Finally, a normalization is introduced,
	
	\begin{equation}
\label{eqn:normalized-soap}
		K\left(\rho_\text{mol}, \rho_\text{mol}'; n, \zeta\right)
		=\left(
		\frac{k\left(\rho_\text{mol}, \rho_\text{mol}';n\right)}
		{\sqrt{k(\rho_\text{mol}, \rho_\text{mol}; n) k\left(\rho_\text{mol}', \rho_\text{mol}';n\right)}}
		\right)^{\zeta}
	\end{equation}
	
	where each density depends on the width $\alpha$ of the Gaussians and on the positions of the atoms, $\{\mathbf{R}_I\}$.
	
	Due to the quadratic scaling of the integrals (for each pair of atoms in the scaffold), it can become inefficient for larger molecules. This is certainly true for very large molecules. 
A remedy for the quadratic scaling is proposed that involves an approximation in terms of an expansion using radial basis functions. 
However, no systematic study has ever scrutinized this radial-basis approximation versus the analytic solution, apart from an initial discussion in the original paper\cite{bartok2013}. 
	A further analysis of the accuracy of the power spectrum or engineered adversarial problems is beyond the scope of this analysis.

	\subsection{Electron Density-Based Comparison: SOED}
	\label{sec:variant}
	
	Instead of attaching a Gaussian function to each atomic position to
    introduce a fuzzy atomic core as a component of a molecular scaffold
    density as in Eq. (\ref{eqn:mol-dens}), one is tempted to exploit the
    electron density of a molecule for the assessment of molecular similarity.
    Not only does this quantity relate to the electronic energy through the
    Hohenberg--Kohn
theorem\cite{hohenberg1964}, but it is also an observable that is accessible in diffraction experiments and from any quantum chemical method. 
Moreover, it also includes a representation of the atomic cores because its maxima indicates the nuclear positions (and even the nuclear charge by virtue of the Kato cusp condition). In addition to this information about the molecular scaffold, the electron density encodes information about the electronic wave function in the valence regions---although their peculiarities are in the tiny details and might require derived fields (such as the Laplacian\cite{popelier2000}) to be clearly visible. 
	
	In this context, it is important to emphasize that the electron density is
    typically calculated from absolute squares of MOs that are decomposed into atomic orbitals (AOs) centered on the atomic nuclei. These AOs can then be represented by standard Gaussian functions available from a basis set library. In this regard, to employ the electron density is even on a technical level very similar to the SOAP scheme based on the molecular scaffold density---although the derivation
will be far more difficult as we will see in the following.
	Hence, the electron density could be taken as a replacement for the molecular scaffold density to represent structural information as does SOAP (best seen with width parameters, $\alpha_I$, that are different for every atomic core),
but now also to encode electronic information. 
	
	Already the simplest electronic structure model will allow us to combine
    the AOs located at the nuclear positions linearly to yield MOs and to then
    yield the electron density, in complete analogy to Eq.
    (\ref{eqn:mol-dens}). However, note that the MOs themselves cannot be used
    to replace the electron density or the molecular scaffold density because
    the coefficients in the linear combination of atomic orbitals (LCAO)
    expansion can take negative values and, in a  superposition of MOs, these
    MO coefficients would cancel each other and all information about
    individual MOs will be lost. 
    Obviously, this problem does not occur for the electron density that is taken
    as a weighted sum of the absolute squares of the MOs (see below).
	
	If the additional electronic structure information encoded in the electron
    density can be harnessed, it might be better suited as a descriptor than SOAP, which encodes only the molecular scaffold. 
A similar reasoning has led Carb\'{o}\cite{carbo1980} in the context of cheminformatics to develop a similarity measure based on the overlap of electron density, which is sometimes called the Carb\'{o} index $r_{A B}$,
	\begin{equation}
r_{A B}=\frac{\int_{V} \rho_{A} \rho_{B} d {\bf r} }{\left(\int_{V} \rho_{A}^{2} d {\bf r}\right)^{1 / 2}\left(\int_{V} \rho_{B}^{2} d {\bf r}\right)^{1 / 2} }
	\end{equation}
and therefore reminiscent of Eq. (\ref{eqn:normalized-soap}) in the SOAP formalism. The derivation, where the first steps are similar to ours, is presented in the appendix of Ref. \citenum{carbo1980} but leaves out the key step of density rotations introduced in SOAP. 
The optimum overlap of $r_{AB}$ is numerically calculated, whereas we aim at an analytical solution.
However, bringing electron densities in a rotatable form in a single-center picture leads to 
significant mathematical overhead as we shall see in the following. 
	
	Because we assume that we will always have results for a simple electronic
    structure model available from which we may take test data, we consider a
    single Slater determinant model. For such an ansatz, which is the basis of
    Kohn--Sham DFT, Hartree--Fock theory, and any approximate Hartree--Fock model, the electron density is the square of $w$ MOs, $\phi_i(\mathbf{r})$, where $\mathbf{r}$ is the coordinate of the electron, 
	
	\begin{equation}
		\rho_\text{el}(\mathbf{r})
		=\sum_{i=1}^{w} 
		n_i
		\left|
		\phi_i(\mathbf{r})\right|^{2}
		\label{eqn:dens}
	\end{equation}
with occupation numbers $n_i$ (that may be taken to be 1 in an unrestricted framework, where $w$ will then be identical to the
number of electrons $N$).
	The MOs are usually expanded as a LCAO, 
	\begin{equation}
		\phi_{i}(\mathbf{r})
		= \sum_A^M \sum_{\mu @ A} c^{(i)}_{\mu} \chi_{\mu} (\mathbf{r}) \ ,
		\label{eqn:lcao}
	\end{equation}
        as in Eq. (\ref{eqn:simple-expansion}) but now with an explicit notion
        of the atom (``$A$'') on which a function $ \chi_{\mu}$ is centered. In
        other words, the above expression explicitly introduces a sum over
        these atomic centers of the basis functions, which we may denote with the German word ``Aufpunkt'' in order to introduce a notion that allows for centers that are not identical with nuclear positions.
	Hence, there is still a total of $m$ basis functions (i.e., the AOs in this
    LCAO), $\chi_\mu(\mathbf{r})$, and their corresponding coefficients are $c^{(i)}_{\mu}$. Gaussian orbitals (GTOs) employed as AO are defined as 
	
	\begin{equation}
		\chi_\mu(\mathbf{r}_{A})=
		N_{\mu}
		f_\mu(\mathbf{r}_{A})
		g_\mu(\mathbf{r}_{A})
		\label{eqn:gto}
	\end{equation}
	
	with the polynomial
	
	\begin{equation}
		f_\mu(\mathbf{r}_{A})
		=\left|\mathbf{r}_{A} \right|^{L_\mu} 
		Y_{L_\mu M_\mu} \left(\Omega_{\mathbf{r}_A}\right) \ ,
		\label{eqn:poly1}
	\end{equation}
	
	the Gaussian
	
	\begin{equation}
		g_\mu(\mathbf{r}_{A})
		=\exp \left(-\zeta_\mu \left|\mathbf{r}_{A}\right|^{2} \right) \ ,
	\end{equation}
	
	and with the normalization constant $N_{\mu}$. Furthermore, we have the local vector from nucleus $A$ at $\mathbf{R}_{A}$ to the field variable $\mathbf{r}$ (see Figure \ref{fig:multicenter}), that is, $\mathbf{r}_{A} = \vert \mathbf{r} - \mathbf{R}_{A} \vert$, the spherical harmonic, $Y_{L_\mu M_\mu} (\Omega_{\mathbf{r}_A})$, which is also centered at the position of nucleus $A$ and hence in the local coordinate system $\mathbf{e}'_{x_A}$, $\mathbf{e}'_{y_A}$, and $\mathbf{e}'_{z_A}$, with orbital quantum number (degree) $L_\mu$ and magnetic quantum number (order) $M_\mu$ and polar angles $\Omega_{\mathbf{r}_A} = (\theta_{\mathbf{r}_A}, \phi_{\mathbf{r}_A})$ with respect to nucleus at $\mathbf{R}_{A}$ and effective nuclear charge $\zeta_\mu$. In the following, we omit the variable dependence for the sake of brevity, that is, $\chi_\mu \equiv \chi_\mu(\mathbf{r})$. As before, we will have to bring this multi-centered approach into a framework where all angles 
are to be taken with respect to a single center, which facilitates the rotation operation and which is taken
to be the center of mass.

	Substituting the AO expansion, Eq. (\ref{eqn:lcao}), into the density expression, Eq. (\ref{eqn:dens}), we obtain
	
	\begin{equation}
		\rho_\text{el}(\mathbf{r}) = 
		\sum_{i}^{w}
		\sum_{A, B}^{M}
		\sum_{\mu @ A \atop \nu @ B}
		n_i
		c^{(i)}_\mu c^{(i)}_\nu
		\chi_\mu \chi_\nu
		\label{eqn:expansion}
	\end{equation}
	
	with basis functions (i.e., ``AOs'') $\chi_\mu$ and $\chi_\nu$ and expansion coefficients  $c^{(i)}_\mu$ and $c^{(i)}_\nu$. As usual, the introduction of a density matrix,
	\begin{equation}
		D_{\mu\nu}
		= \sum_{i}^{w}
		n_i c^{(i)}_\mu c^{(i)}_\nu
	\end{equation}
	(which is not to be confused with the Wigner D-matrices) brings the expression for the electron density into a more compact form,
	
	\begin{equation}
		\rho_\text{el}(\mathbf{r}) = 
		\sum_{A, B}^{M}
		\sum_{\mu @ A \atop \nu @ B}
		D_{\mu\nu}
		\chi_\mu \chi_\nu
		\ .
		\label{eqn:D-dens}
	\end{equation}
Moreover, for notational convenience, we introduce a product GTO,
	
	\begin{equation}
		\begin{split}
			\chi_{\mu\nu}(\mathbf{r}_A,\mathbf{r}_B)
			= N_{\mu\nu} 
			f_\mu(\mathbf{r}_{A})
			f_\nu(\mathbf{r}_{B})
			g_{\mu}(\mathbf{r}_{A})
			g_{\nu} (\mathbf{r}_{B})
		\end{split}
	\end{equation}
	
	with the normalization constant $N_{\mu\nu} = N_{\mu}N_{\nu}$

	Because we want to rotate one of the electron densities with respect to the other, the expression must be made dependent on the rotation angles. However, as in the case of SOAP, the angles in the spherical harmonics
are defined with respect to the local coordinate system at a nucleus but must be defined with respect to a single center that 
will be the center of mass. 
For this, we follow the derivation of Kaufmann and Baumeister\cite{kaufmann1989} here.
	However, the result from the exact derivation, which we sketch in the supporting information,
 is very long-winded and turns out to be computationally costly to evaluate.
	Therefore, we propose two ways of approximating the GTO in order to simplify the expression:
        First, we neglect all products of higher order spherical harmonics;
        for example, no product of two $p$ functions was considered, if they are positioned at different nuclei.
        Although this simplifies the derivation, it is still very involved. 
        Therefore, we eliminate the polynomial factors by replacing all $p$ basis functions
        with lobe functions, that is, $s$-type functions with shifted centers that resemble functions of higher angular
       momentum quantum number (see supporting information for details on how these new Aufpunkte were determined by starting
       from the nuclear positions). This is also possible for higher-orbital-momentum functions such as
       $d$ functions.
       In this way, we generate an all-$s$-type
        basis set that sufficiently represents the electron density and for which the calculated MO coefficients
        can be inherited. Note that, from a technical point of view, the new Aufpunkte of these lobe functions are then to be treated as (new artificial) nuclear positions
        in the evaluation procedure for SOAP, and therefore, SOED can now be
        evaluated with a SOAP-type procedure. 

By virtue of the Gaussian product theorem, each product of the two Gaussians $\chi_\mu$ and $\chi_\nu$ 
in Eq. (\ref{eqn:D-dens}) becomes another Gaussian.
Hence, we can collapse the double indices $\mu\nu$ 
in Eq. (\ref{eqn:D-dens}) into a single index, which we call $\check{I}$ in order
to associate it to the corresponding expression in the SOAP derivation: because we consider a
lobe-basis of $s$-functions only, we obtain the electron density as an expansion into $s$-functions
only with Aufpunkte given at positions denoted by $\check{I}$, some of which are actual nuclear positions:
\begin{equation}
\rho_{\mathrm{el}}(\mathbf{r})
=\sum_{\check{I}}^{\check{m}} D_{\check{I}} \chi_{\check{I}}
\end{equation}
All of these positions $\check{I}$ are subject to the SOAP rotation and overlap procedures and can be
treated like ghost atoms.
Hence, we have recovered a generalized version of SOAP with weights $D_{\check{I}}$ in front of the basis functions
$\check{I}$, each of which will, in general, carry a different exponent (rather than a common exponent $\alpha$,
i.e. the standard choice for SOAP). Recall that the weights $D_{\check{I}}$ contain products of MO coefficients
and occupation numbers (or first-order reduced density matrices), which can be determined in an electronic
structure calculation (e.g., in a Hartree-Fock calculation). Of course, this will require one storing electronic
structure data in addition to the nuclear coordinates of a molecule, but that should not present a hurdle as
the molecular structures will typically be optimized with an electronic structure method whose wave-function
ingredients then simply need to accompany the Cartesian coordinates in a data base in order to be exploited by
a machine learning ansatz based on SOED.

Finally, we obtain, by virtue of the direct analogy with SOAP, 
for the overlap of two electron densities,
\begin{equation}
 \label{eqn:k-soed}
k\left(\rho_{\mathrm{el}}, \rho_{\mathrm{el}}^{\prime} ; 2\right) = 
\sum_{\check{I}, \check{I}^{\prime}}^{\check{m},\check{m}'} \sum_{l, m, m^{\prime}} 
\frac{D^2_{\check{I}} D^2_{\check{I}^{\prime}}}{2 l+1} \tilde{I}_{m m^{\prime}}^{l *} \tilde{I}_{m m^{\prime}}^{l}
\end{equation}

\subsection{Numerical Comparison of SOAP and SOED}
	
\begin{figure}[h]
	\includegraphics[width=0.99\textwidth]{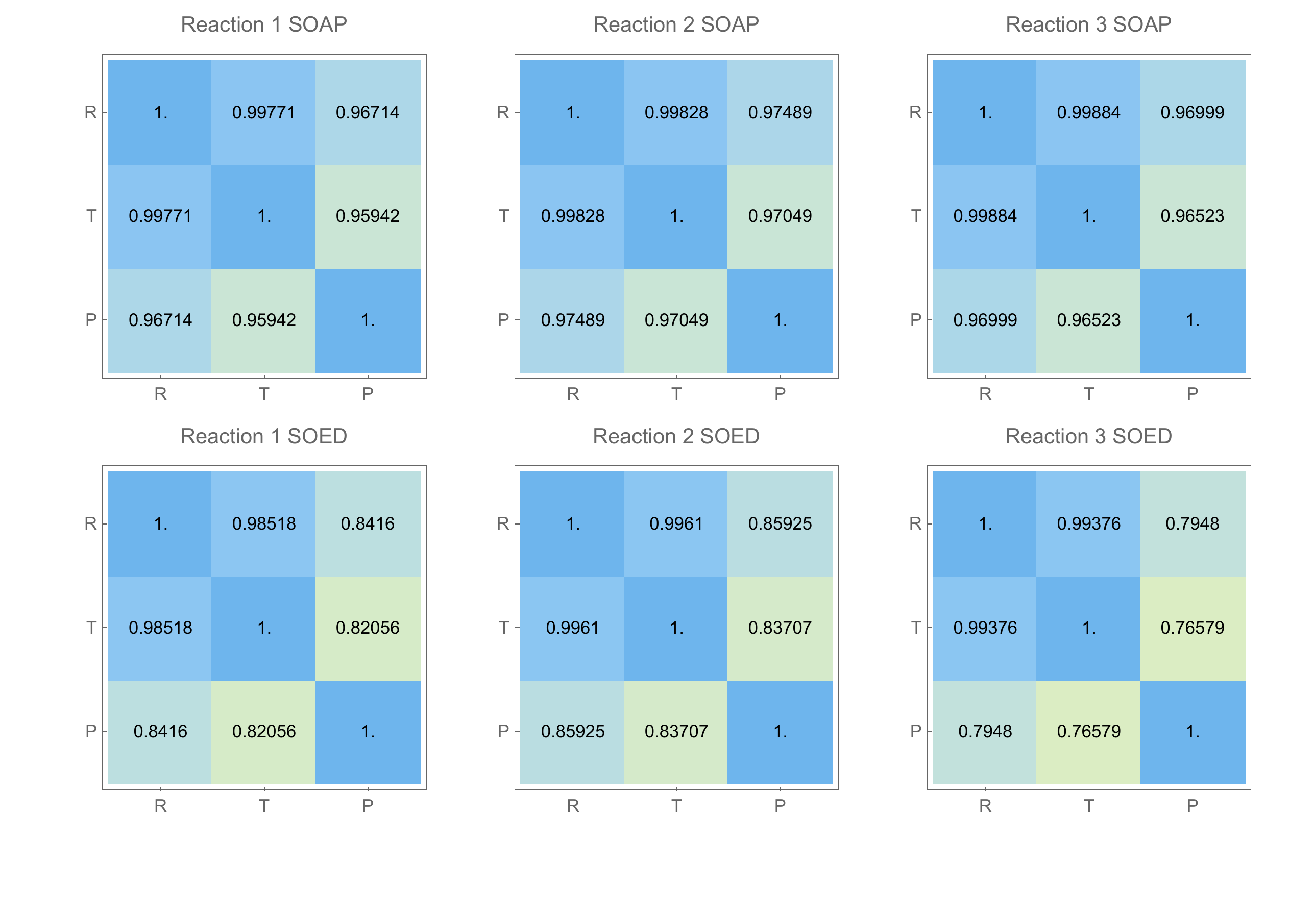}
	\caption{Comparison of SOAP (top panel) and SOED (bottom panel) for the
    three test reactions; ``R'' denotes the reactant, ``T'' the transitions
    state, and ``P'' the product. 
		}
	\label{fig:soap-soed}
\end{figure}

	We show in Figure \ref{fig:soap-soed} a comparison for the SOAP and SOED
    similarity results for the reactions in Figure \ref{fig:rxn}.
    On both axes are the reactant (R), the TS (T), and the product (P) of the reaction.
	The diagonal is 1, as self-similarity is perfect by design.
    The first row shows SOAP and the second row shows the SOED results.
	As one can see, the electron density based descriptor is more sensitive to the reaction progress, because it drops from reactant
	to TS more than traditional SOAP. 
	Again, the reactant and the TS are more similar in all descriptors shown than TS and product, implying an early TS again, as we already 
	found for the CM and the energy diagram.
However, product and TS are less similar than product and reactant, which seems unexpected at first sight. However,
we note that this is actually reasonable if a purely structure-based descriptor that is largely independent of directional
information (owing to the radial Gaussians involved in SOAP) favors similarity in terms of equilibrium
bond lengths over similarity between stretched and equilibrium bond lengths in one constitutional isomer.

\begin{figure}[h]
    \includegraphics[width=0.99\textwidth]{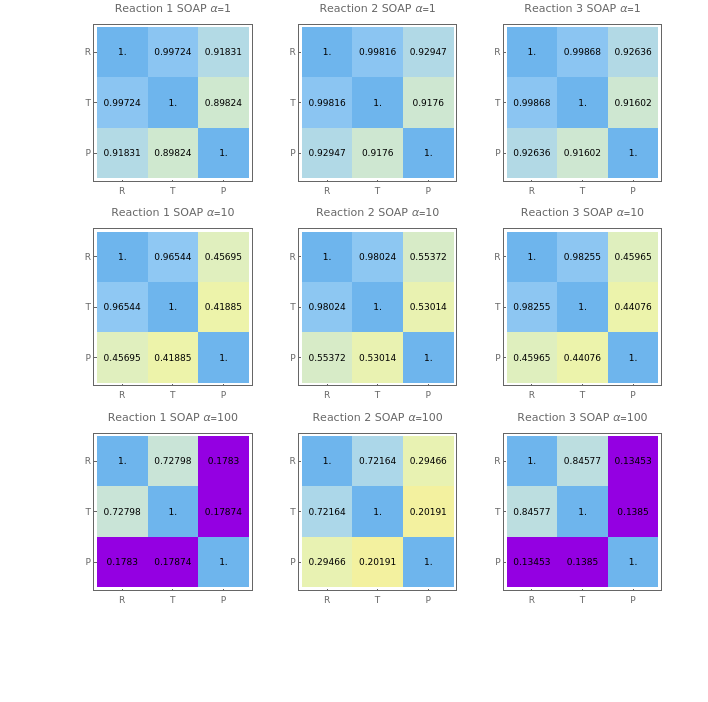}
	\caption{Comparison of SOAP for three different values of the common
    exponent $\alpha$ for the three test reactions; ``R'' denotes the reactant,
    ``T'' the transitions state, and ``P'' the product. 
	}
	\label{fig:soap-alphas}
\end{figure}

One reason for the larger dissimilarity obtained for the SOED kernel is the fact that the parameter in the exponent of the
Gaussian functions is no longer a hyperparameter. By contrast to SOAP with a
fixed value $\alpha$ for all nuclei, the
analogous parameter in a Gaussian basis set that describes the AOs is fixed to represent these orbitals.
In order to demonstrate how sensitive SOAP is in this respect, 
a comparison of SOAP results is shown in Figure \ref{fig:soap-alphas}
that was obtained for the three reactions 
with varying $\alpha$: $\alpha=1$, $\alpha=10$, and $\alpha=100$
(recall that the standard value is $\alpha=0.4$). 
It is apparent from Figure \ref{fig:soap-alphas} that when rotating one SOAP scaffold density with respect to the other, narrower peaks will have less overlap compared to when they are spread out.

\clearpage

\section{Comparison of Molecular Similarity Descriptors}

On the basis of the three reactions studied in this work, we may 
draw some conclusions from a comparison of all descriptors considered here.
The CL and the eigenvalues of the CM
deliver similar results for the elementary steps considered.  Yet, the
former may be preferred over the latter as it retains interpretable meaning through a
direct and unique atomic assignment of parts of the external potential.
Furthermore it is cheaper to calculate, as no diagonalization step is needed. 
Moreover, we emphasize that by comparison to the eigenvalues of the CM
the CL entries allow one to track
the progress of a chemical reaction easily and classify the TS structure as early or late,
as shown in Figure 6 above.

\begin{figure}[h]
    \centering
        \includegraphics[width=\textwidth]{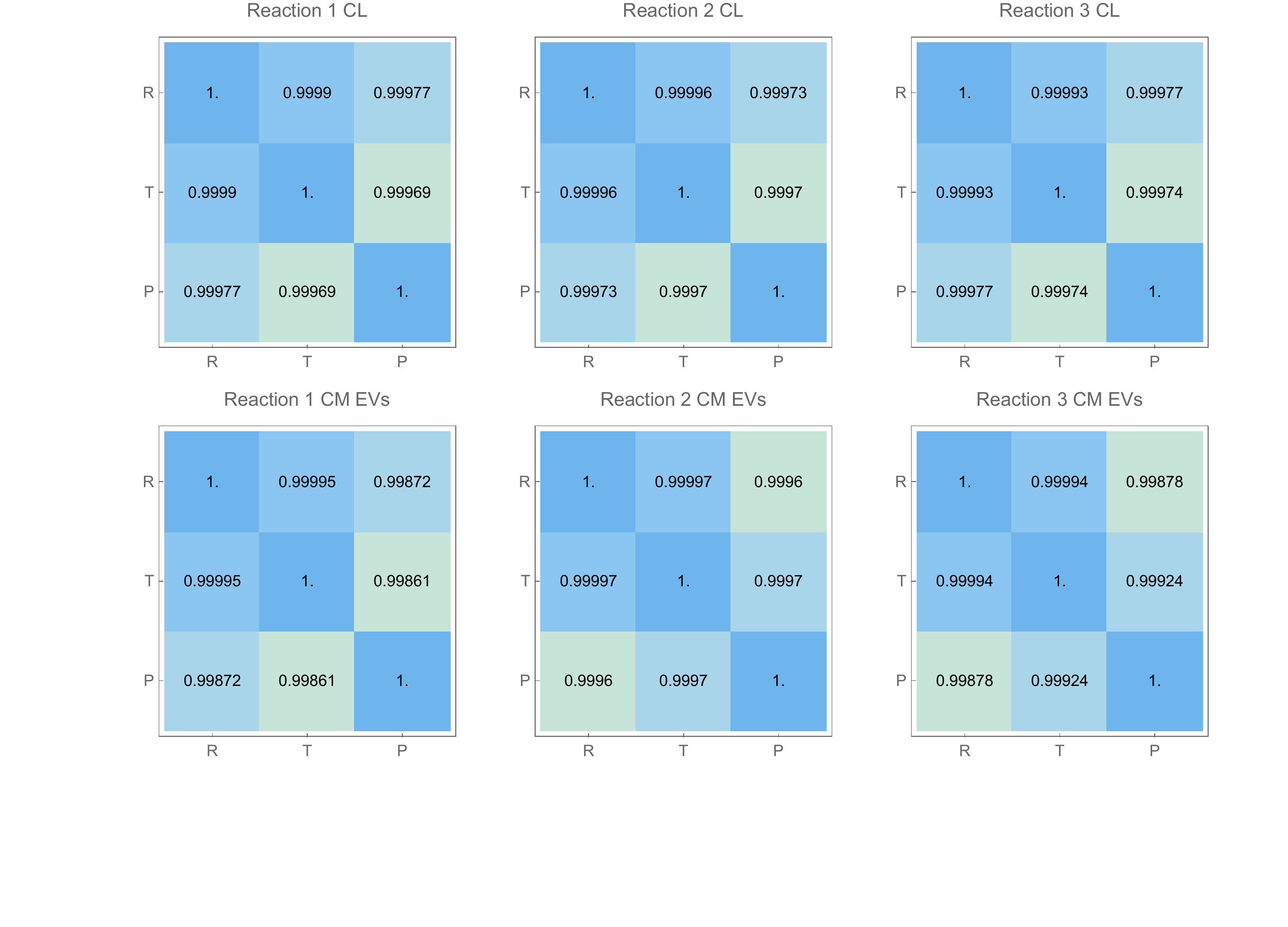}
        \caption{Comparison of CL (top panel) and eigenvalues of the
    CM (EVs, bottom panel) for the three
    reactions 1-3; ``R'' denotes the reactant, ``T'' the transitions state, and ``P'' the product.
}
        \label{sfig:cl-cme}
\end{figure}

Figure \ref{sfig:cl-cme} presents a comparison of the eigenvalues of the CM 
and the CL for reactions 1--3 of Figure 1, in analogy to the figures created above for the
comparison of SOAP and SOED. As in the case of SOAP and SOED, the diagonal elements all equal 1 
because self-similarity is perfect by design. The first row shows the CL and the second row shows the
CM eigenvalue results.
The eigenvalues of the CM appear slightly more sensitive to the
reaction progress.
As for SOAP and SOED, the reactant and the TS are more similar in most descriptor values 
than TS and product, implying an early TS.
Moreover, product and TS are less similar than product and reactant.
However, all data points in Figure \ref{sfig:cl-cme} are very similar, and there is no natural way to
enhance the differentiation of the different structures (in sharp contrast to SOAP; see also below).

Compared to the Coulomb-type descriptors, SOAP is far more expensive to evaluate in terms of the
computational effort, and obviously, SOED requires even more computational resources.
As already mentioned in the original publication\cite{bartok2013}, the evaluation
for the integrals can be very time consuming, because $I_{mm\prime}^l$ must
be evaluated for every neighboring pair, $I$ and $I'$. Accordingly, this can represent a
bottleneck for big-data high-throughput machine learning applications. Depending on how accuarate the numerical
expansion in terms of spherical harmonics needs to be (variables $m$, $m'$, and $l$)
the scaling can be very unfavorable for SOAP---and also for SOED, which relies on the same framework.

Because SOED requires a number of Gaussian functions equal to the number of one-electron basis
functions in a given quantum chemical calculation, the number of neighbors is
much higher than the number of atoms. In addition, higher angular momentum functions create
an increase of the number of basis functions (and hence neighbors) needed because of the lobe functions 
(each $p$-function is approximated with 6 $s$-functions).
Because the scaling of the calculation $I_{mm\prime}^l$ is quadratic, more basis
functions will be computationally much more expensive. 

A way out of this dilemma might be a further reduction of basis
functions or even a complete replacement of the electronic density by an approximation with a
molecular scaffold density but set up with different exponents $\alpha_I$ for
different atoms $I$ in such a way that the electronic density is still as well approximated as possible.
In this context, we recall that Figure 9
demonstrated the role of the magnitude of these exponents, which can allow for a better differentiation of structures
along the reaction coordinate.

The fact that SOED requires a quantum chemical calculation to produce the orbital and hence the density information
might become another bottleneck. However, in the case of high-throughput virtual screenings and automated reaction
mechanisms explorations, such data will be available without additional costs.

While comparing the Coulomb interaction-based descriptors and the
density-based descriptors in terms of predictive performance needs to be
evaluated on a large data set in future work, we can already draw qualitative conclusions from the
analysis of the elementary step. As we can see in Figure
\ref{fig:soap-alphas} for SOAP and SOED and analogously for the CL
and CM in Figure \ref{sfig:cl-cme}, the TS structure is always 
closest to the reactant. In the case of reactions 2 and 3 for the eigenvalues of
the CM, the TS structure is even more similar to both the reactant
and the product than they are similar to each other.  Thus, despite the inclusion
of the coordinates and nuclear charges that are formally sufficient to
describe the electronic structure of a molecule, it appears that both types of 
descriptors mainly represent the geometric structure and ignore important
electronic effects, leaving the TS similarity paradox unresolved.
This points to the need to encode further information about the electronic structure
of molecular structures into similarity descriptors or fingerprints
in order to discriminate between the different electron-correlation regimes of
stable intermediates and TS structures (where chemical bonds are being broken or formed).

\section{Conclusions}

In this work, we considered an elementary expression for the electronic energy that allowed
us to discuss two widely used descriptors of molecular similarity in machine learning from the point of view of electronic
structure theory:
CM and SOAP. 
We showed how to ground their definitions into electronic structure theory by (i) introducing CL that allowed us
to scrutinize the rather arbitrary diagonal entries of the CM and its non-transparent diagonalization step and by (ii)
relating the fuzzy density that encodes molecular structure for SOAP to the actual electron density, which then also carries
electronic structure information directly into the descriptor. 

Our formal discussion was accompanied by a single example that served the purpose to illustrate the results one
obtains with the standard descriptors CM and SOAP and with our new descriptors CL
and SOED. The single example was chosen to provide structures connected through an elementary reaction step---reactant, TS,
product---that is, structures that are rather similar by definition and that occur in chemical reaction networks. 
Although the few data points are far from a big-data machine learning approach, our example allowed us to study directly the variation in the descriptors along a coordinate that connects the three 
types of structures. Whereas structural change is therefore continuous, the electronic structure is different, which
is the reason why the TS structure acquires a higher energy than the reactant and product structures.
This situation therefore introduces a peculiar twist that would allow one to argue that the TS structure
is more similar to either product and reactant than product and reactant are similar to one another, while the difference
in electronic structure prompts one to argue that the stable intermediates, that is, product and reactant, should be more
similar to one another with respect to the nature of their electronic structures.

For the CL, we found that the $\mathcal{F}_{\text{n}}^{(J)}$ descriptor yields interpretable traces
by contrast to the convoluted eigenvalues of the CM. At the same time, it contains the same
information and can therefore replace the CM eigenvalue features, also alleviating the need for the diagonalization step.

We emphasize that traces of these features along reaction coordinates clearly correlated with the change in 
molecular structure and are able to identify the TS structure in the feature.
Hence, the electronic difference of the TS compared to the stable intermediates (reactant and product) 
will be detectable in these features, if they are considered relative to one another.

We found that SOED is more sensitive than SOAP with the standard $\alpha$
parameter. This is a consequence of the widths of the Gaussians that are present
in the representation of the electron density. We
found that narrowing the Gaussians that compose the SOAP kernel also increases sensitivity, so that
SOAP, which is much easier to evaluate than SOED, could be used instead, but with a parameter $\alpha$
in the exponent that is larger than the standard one.

  By analyzing an elementary step through the lenses of the CM
  and SOAP, we were able to show their connection to the first principles of
  quantum mechanics. The dependence on the nuclear charge, explicitly in the
  case of the CM and implicitly through the electron density in SOED (and through its molecular scaffold approximation in SOAP), makes them
  conceptually suitable for the whole of chemical space.

  In this quantum chemical study, we studied single elementary steps to
  draw detailed conclusions on individual changes of the descriptors. Clearly,
  a big data approach needs to be taken to demonstrate actual usefulness in
  machine learning applications.
  In future work, we will build upon our findings and elaborate on molecular
  similarity in the context of a huge number of elementary steps (on the same and on different Born--Oppenheimer surfaces). Only such work will eventually allow us to rate the value of our new descriptors for machine learning purposes compared to known descriptors.

\section*{Supporting Information}
Further CL analysis, details on the derivation and implementation of SOED, Cartesian coordinates of all
molecules, and additional graphical representations are provided as additional material in the Supporting Information.

	\section*{Acknowledgements}
Financial support by the Swiss National Science Foundation through project no. 200021\_182400
is gratefully acknowledged.

	\section*{Appendix: Computational Methodology}
	The quantum chemical calculations for the model reactions were carried out with Orca 5.\cite{neese2020}
	We performed unrestricted Kohn--Sham PBE\cite{perdew1996a} structure
    optimizations with the SVP basis set with density
    fitting.\cite{weigend2006} Our molecular similarity descriptors were then obtained from Hartree--Fock single-point calculations.
The atomic descriptors $\mathcal{F}^{(J)}_\mathrm{n}$ and $\mathcal{F}^{(J)}_\mathrm{e}$ were obtained with the program  
Serenity\cite{unsleber2018,barton2020}, which calculated the nuclear attraction integrals contracted with the density matrix elements and the nucleus--electron interaction with
unrestricted Kohn--Sham PBE\cite{perdew1996a} and the def2-SVP basis set.\cite{weigend2005}

	For SOED, we obtained the MO coefficients for the electron density in Hartree--Fock calculations with the tiny STO-3G basis set 	
	\cite{hehre1969} with Gaussian\cite{g16}, where the keyword \texttt{Integral(SplitSP)} had to be 
	applied to not obtain \texttt{S=P} contracted orbitals but regular $s$ and $p$ orbitals.
        We note that despite its small size, this basis set already produces the main features of the electron density (non-isotropic local effects through minimal polarization by basis functions on atomic neighbors and an element-specific maximum of the density distribution at the various atomic nuclei) that makes SOED different from SOAP.

We implemented Eq.\ (\ref{eqn:kernel1}), the SOAP kernel, in Mathematica\cite{Mathematica} and this code is available from the authors.
In addition, we established a python implementation for which we used
NumPy\cite{harris2020} array programming (vectorization). The data structure
\texttt{ndarray} harnesses the CPU's SIMD (single instruction, multiple data)
architecture for a significant speed-up compared to a loop-based
implementation. Furthermore, we parallelized the calculation with the library
\texttt{Ray}.\cite{moritz2018}
Our implementation is available open source.\cite{gugler2022a}

According to Eq.\ (\ref{eqn:soap-overlap}), the overlap for the SOAP procedure
is dependent on an infinite sum that originates from the Rayleigh expansion in
Eq.\ (\ref{eqn:rayleigh}) and the rotation of the spherical harmonics as an
expansion of Wigner-D matrices in Eq.\ (\ref{eqn:wigner-rotation}), which
converges uniformly. Even though we avoid the standard power spectrum
approximation, which would introduce another approximation, we can
approach the exact overlap only within numerical accuracy. As the authors of the original paper\cite{bartok2013} noted, the pairwise evaluation for all $I$ and $I'$
leads to a big computational overhead. Because we approximated the $p$-functions by (lobe) $s$-functions, 
many terms are created in this double sum. Our largest molecule from reaction 3 in Figure \ref{fig:rxn} 
then requires 333 $s$-functions. For this reason, it is hard to reach a high $l$ in the expansion, as $m$ and $m'$ 
range from $-l$ to $l$ in steps of one. Hence, for the results presented in Figure \ref{fig:soap-soed}, the maximum
value for $l$ was 3 for SOED and 5 for SOAP.

In the original reference introducing SOAP\cite{bartok2013}, it had already been noted that
the computation of SOAP can be expensive, 
because the terms inside the sums of Eq.\ \ref{eqn:final-ints} have to be evaluated for each pairwise
interaction of atoms. The sums over $l$, $m$, and $m'$ scale with $O(L_\mathrm{max}^3)$, where
$L_\mathrm{max}$ is the truncation of the infinite sum over $l$, yielding an overall complexity of
$O(M^2 L_\mathrm{max}^3)$. In SOED, Eq.\ (\ref{eqn:k-soed}), where the density is based on $m^2$ basis functions for
each density, the evaluation becomes even more expensive as $O(m^4 L_\mathrm{max}^3)$, where
the number of basis functions, $m$, is in all practical cases much bigger than the number of atoms, $M$.
The straightforward way to treat this unfavorable scaling is to trade off accuracy for speed and reduce $L_\mathrm{max}$.
For all but the smallest molecules, larger basis sets than those on the order of STO-3G will likely not
be feasible. As in SOAP with the power spectrum, it might be possible to simplify the nested sum with 
some mathematical transformations to gain a speed-up. 


\end{document}